\documentclass[smallextended]{svjour3}

\usepackage[utf8]{inputenc}
\usepackage{mathpazo}
\usepackage{amsmath}
\usepackage{amssymb}
\usepackage{standalone}
\usepackage{graphicx,wrapfig,lipsum}
\usepackage{mathtools}
\usepackage{bigints}
\usepackage[style=numeric,minbibnames=3,maxbibnames=3,autopunct=true,babel=hyphen,hyperref=true,abbreviate=false,backend=biber]{biblatex}
\usepackage{hyperref}
\usepackage{epstopdf}
\usepackage[utf8]{inputenc}
\usepackage[english]{babel}
\usepackage{caption}
\usepackage{epstopdf}
\usepackage{float}
\usepackage{multicol}
\usepackage{lipsum}
\usepackage[normalem]{ulem}
\usepackage[rgb,dvipsnames]{xcolor}
\usepackage{tikz}
\usepackage{bm}
\usepackage{subcaption}
\usepackage{booktabs}
\usepackage[margin=1in]{geometry}
\usepackage[math]{cellspace}
\AtBeginDocument{
    \heavyrulewidth=.08em
    \lightrulewidth=.05em
    \cmidrulewidth=.03em
    \belowrulesep=.65ex
    \belowbottomsep=0pt
    \aboverulesep=.4ex
    \abovetopsep=0pt
    \cmidrulesep=\doublerulesep
    \cmidrulekern=.5em
    \defaultaddspace=.5em
}

\definecolor{MyBlue}{rgb}  {0.1,0.1,0.9}
\definecolor{MyRed}{rgb}   {0.9,0.1,0.1}
\definecolor{MyGreen}{rgb} {0.05,0.4,0.05}
\definecolor{burntorange}{rgb}{0.8, 0.33, 0.0}
\definecolor{NeilMagenta}{rgb}{0.8, 0.1, 0.8} 

\newcommand \bl{\color{black}}
\newcommand \rd{\color{black}}

\newcommand{\diff}[2]{\frac{\mathrm{d} #1}{\mathrm{d} #2}}
\newcommand{\pdiff}[2]{\frac{\partial #1}{\partial #2}}

\newcommand \beq{\begin{eqnarray}}
\newcommand \eeq{\end{eqnarray}}
\newcommand \beqno{\begin{eqnarray*}}
\newcommand \eeqno{\end{eqnarray*}}
\newcommand \bit{\begin{itemize}}
\newcommand \eit{\end{itemize}}

\newcommand{\F}{\mathcal{F}}
\newcommand{\G}{\mathcal{G}}
\newcommand{\D}{\mathcal{D}}

\begin{document}
\title{Fixed and Distributed Gene Expression Time Delays in Reaction-Diffusion Systems}
\titlerunning{Fixed and Distributed Gene Expression Time Delays in Reaction-Diffusion Systems}
\author{Alec Sargood \and  Eamonn A. Gaffney \and Andrew L. Krause }
\authorrunning{A. Sargood \and E. A. Gaffney \and A. L. Krause}

\institute{A. Sargood \and E. A. Gaffney \and A. L. Krause
    \at Wolfson Centre for Mathematical Biology, Mathematical Institute, University of Oxford, Andrew Wiles Building, Radcliffe Observatory Quarter, Woodstock Road, Oxford, OX2 6GG, United Kingdom \\
    A. L. Krause 
    \at Mathematical Sciences Department, Durham University, Upper Mountjoy Campus, Stockton Rd, Durham DH1 3LE, United Kingdom
    }

\date{Received: date / Accepted: date}

\maketitle

\begin{abstract}
Time delays, modelling the process of intracellular gene expression, have been shown to have important impacts on the dynamics of pattern formation in reaction-diffusion systems. In particular, past work has shown that such time delays can shrink the Turing space, thereby inhibiting patterns from forming across large ranges of parameters. Such delays can also increase the time taken for pattern formation even when Turing instabilities occur. Here we consider reaction-diffusion models incorporating fixed or distributed time delays, modelling the underlying stochastic nature of gene expression dynamics, and analyze these through a systematic linear instability analysis and numerical simulations for several sets of different reaction kinetics. We find that even complicated distribution kernels (skewed Gaussian probability density functions) have little impact on the reaction-diffusion dynamics compared to fixed delays with the same mean delay. We show that the location of the delay terms in the model can lead to changes in the size of the Turing space (increasing or decreasing) as the mean time delay, $\tau$, is increased. We show that the time to pattern formation from a perturbation of the homogeneous steady state scales linearly with $\tau$, and conjecture that this is a general impact of time delay on reaction-diffusion dynamics, independent of the form of the kinetics or location of the delayed terms. Finally we show that while initial and boundary conditions can influence these dynamics, particularly the time-to-pattern, the effects of delay  appear robust under variations of initial and boundary data. Overall our results help clarify the role of gene expression time delays in reaction-diffusion patterning, and suggest clear directions for further work in studying more realistic models of pattern formation. \end{abstract}

\keywords{Pattern formation \and time delay \and linear instability analysis}

\maketitle

\section{Introduction}

The self-organisation of cells into an apparent order appears across many different fields within biology. For example, the distribution of cells during the developmental process of an embryo, the growth of cancerous tissue \cite{morph}, vertebrate limb development \cite{miura1,glimm,miura2}, and pattern formation on animal coats (e.g. spots on a jaguar \cite{painter}, feathers on birds \cite{bailleul}). Wolpert \cite{wolpert} presented the idea that, underpinning the development of shape and form (morphogenesis) is a cell's ability to differentiate according to its position in space and time. Furthermore, the concentration of  signalling molecules  (morphogens), or the concentration gradients of certain morphogens across a spatial domain of cells, affects the cell differentiation mechanism, and thus cells adopt a state relative to the concentration of a specific morphogen that they are exposed to.

The mechanism allowing cells to adopt an appropriate state is known as differential gene expression, and depends crucially on the communication between cells, achieved through cellular signalling \cite{gaffmonk}. Typical reaction-diffusion systems modelling such signalling implicitly assume a negligible timescale on which the internal cellular signalling and gene expression processes occur. The gene expression process, however, is extremely complex and proceeds through several stages \cite{gaffmonk}, including gene transcription and gene translation. These sub-processes can take large amounts of time, and it has been experimentally shown that these time delays are typically on the order of minutes, but in some cases can be as large as several hours \cite{gaffmonk,tennyson}. Such time delays can therefore approach the timescale of pattern formation process itself. For example, the basic body plan of a zebrafish is established in less than 24 hours \cite{kimmel}. Hence it seems important to assess the impact of such time delays on pattern formation in developmental biology.

In 1952, Alan Turing proposed that the pattern formation process could be mathematically modelled on a purely chemical basis via the interaction of morphogens, whose evolutions are described by a system of coupled reaction-diffusion equations \cite{turing}. Turing showed that a stable steady state, robust to small perturbations in the spatially homogeneous setting (i.e.~without diffusion), could become unstable and sensitive to small perturbations with the introduction of diffusion, leading to spatially inhomogeneous patterns. Cell fate decisions are then based on these morphogen concentrations, where regions of high morphogen or low morphogen concentration can lead to different cell fate decisions. Turing's model is therefore one of pre-patterning, where the morphogen pattern concentrations across a spatial domain are modelled, which in turn lead to cell fate decisions at a later stage. Typical reaction-diffusion systems in the context of Turing pattern formation consist of two partial differential equations describing the interaction and evolution of two morphogens, the \textit{activator} and \textit{inhibitor}. Empirical evidence suggests that Turing instabilities are present in real biological systems, and can be used to explain complex biological phenomena \cite{yigaffneyli,molecular,miura,miura2,sick}. However, whether Turing patterns {\rd may  be found} experimentally in biological systems with simple two-species systems is still very much an active field of research \cite{bespoke}.

Time delays have been investigated in the context of Turing patterns, both numerically and analytically, through the incorporation of constant fixed time delays. One of the canonical reaction-diffusion mechanisms that exhibits Turing instabilities is the Schnakenberg model \cite{schnakenberg}, also known as the activator-depleted model. This system has been extensively studied in the context of Turing pattern formation with incorporated gene expression time delays. Two biologically motivated variants incorporating time delays are the ligand-internalisation (LI) and reverse ligand-binding (RLB) models \cite{leegaffney}. The LI model places gene expression time delay in purely the activator dynamics, whereas the RLB model contains time delay in both the activator and inhibitor dynamics.

Numerical results in \cite{gaffmonk} on the LI model showed that the time taken until pattern formation occurs drastically increases as the time delay in the model is increased. In particular, small delays on the order of minutes can cause a large increase in time-to-pattern, on the order of several hours, compared to a model with no time delay. This highlights the impact of including gene expression time delays, especially when considering patterning events that occur on a fast timescale. 

The two papers \cite{jiang, yigaffneyli} consider both the LI and RLB variants of the Schnakenberg model. Using linear analysis and numerical simulations, the results in both papers suggest that the RLB model can exhibit spatially inhomogeneous temporal oscillations, as well as de-stabilisation of spatially inhomogeneous steady states, inhibiting pattern formation via Turing instabilities.  The results in \cite{yigaffneyli}  indicate that extensive ligand internalization may inhibit the formation of patterns within some   regions of the parameter space.  Results in both \cite{leegaffney,leegaffmonk} suggest that time delay causes a significant effect on the time taken until pattern formation occurs.  A final observation from \cite{gaffmonk,leegaffmonk}, for both LI and RLB models, is that an increasing time delay may also increase the sensitivity of the final pattern formed to variations in initial conditions.

In contrast, analysis of one-dimensional spike solutions of the Gierer-Meinhardt (GM) model in \cite{fadai1,fadai2} shows that the placement of delay terms in the model can affect the size of the parameter regimes for which the spike solution is linearly stable. It was found that, depending on the positioning of time delayed terms, an increasing time delay can have a stabilising or de-stabilising effect, enlarging or shrinking the stable parameter region of the spike. Further details of spike solutions of the GM model and their stability analysis can also be found in \cite{spike}. This analysis highlighted the importance of time delay positioning in the GM model for the stability of spike solutions. 

The key insights from the literature on delayed reaction-diffusion models above are that fixed time delays can (a) change the space of parameters in which spatially inhomogeneous steady states are observed (the Turing space), and (b) delay the time for patterns to form. These results have implications on the use of such models as mechanistic explanations of pattern formation in development. Namely, in cases where the Turing space shrinks for realistic sizes of delay, this can challenge the robustness of Turing pattern formation, which is already an open problem in using non-delayed reaction-diffusion models to explain robust biological phenomena \cite{maini2012turing, scholes2019comprehensive}. Additionally, the increasing time needed for patterns to form can become an obstacle in quantitatively relating  these models to developmental perspectives for patterning events that occur on relatively fast timescales \cite{leegaffmonk}. Thus our first objective will be to explore these observations in  variants of Schnakenberg and GM systems in more detail by explicitly constructing Turing spaces via a linear stability analysis, and exploring the time taken for patterns to form as a function of the delay time.

The current literature on Turing pattern formation in development has primarily considered gene expression dynamics via fixed time delays in reaction-diffusion processes. On a cellular level however, the biological processes responsible for gene expression are inherently stochastic \cite{raj,elowitz,mcadams,paulsson}.  Distributed time delays have been incorporated to model biological phenomena such as hematopoiesis, and lactose operon dynamics \cite{newdist}, Wnt/$\beta$-catenin signalling pathways \cite{signal}, and Oncolytic virotherapy treatments for cancer \cite{cancer}, among other applications. A distributed delay can be thought of as a more `general and realistic' \cite{cancer} approach to modelling, on a larger scale, a process which may possess an intrinsic stochasticity on a small scale. Within the context of Turing pattern formation, introducing a fixed time delay into the reaction-diffusion mechanism is a simplification of the underlying biological process on a microscopic level. This leads us to our second objective: to model distributed gene expression time delays at the macroscopic level to ascertain if the results obtained in the case of fixed time delays are sensitive to the shape of a distribution of delays, which is in principle a more realistic model for gene expression dynamics \cite{bratsun,krausenew}. Of particular interest within this objective is to assess whether the distribution of time delays induces substantial changes compared to a Dirac-delta kernel for the time delay, noting  the latter is equivalent to a fixed time delay.

In Section \ref{Model_Sect} we outline a general formulation of reaction-diffusion systems with distributed time delay, as well as some specific reaction kinetics that we will examine. We briefly review the underlying theory of Turing pattern formation and carry out a linear stability analysis of our general distributed delay model in Section \ref{Turing_Analysis_Sect}. In Section \ref{Turing_Application_Sect}, we apply this instability analysis to explore how Turing spaces change for increasing time delays in fixed and distributed scenarios. We compare these predictions from the linear theory against numerical solutions in Section \ref{Numerical_Sect}, as well as exploring questions of the time-to-pattern as a function of delay, boundary, and initial conditions. We close with a discussion of our results in Section \ref{Discussion_Sect}, further highlighting the importance of the placement of time delayed terms within the models, and noting the surprising fact that distributed time delays have almost no influence on pattern formation dynamics compared to fixed delay models with the same mean delay.

\section{Models of Gene Expression Time Delay}\label{Model_Sect}

We consider two morphogens, $u$ and $v$, obeying the following non-dimensional reaction-diffusion equations:
\begin{equation}\label{genform}
    \begin{split}
        \pdiff{u}{t}&=\frac{\epsilon^2}{L^2}\pdiff{^2u}{x^2}+f(u,v,\F(u,v)),\\
        \pdiff{v}{t}&=\frac{1}{L^2}\pdiff{^2v}{x^2}+g(u,v,\G(u,v)),
    \end{split}
\end{equation}
where $x\in\Omega=[0,1]$, $t>0$ and 
$$
\F(u,v)=\int_{\tau_1}^{\tau_2}K(s;\bm{p})F(u_s,v_s)\ \text{d}s, \quad \G(u,v)=\int_{\tau_1}^{\tau_2}K(s;\bm{p})G(u_s,v_s)\ \text{d}s,
$$
with $u_s=u(x,t-s)$ and $v_s=v(x,t-s)$ the function evaluations of $u$ and $v$ at some time delay $s$, which is an integration variable.
We also have that $\tau_1$ and $\tau_2$ are the minimum and maximum time delays of the distributions, $L^2$ is the non-dimensional  scaling of the domain length, and $\epsilon^2$ the diffusion ratio between the activator $u$ and inhibitor $v$.  Unless otherwise stated, we use the same value of $\epsilon^2=0.001$ as in \cite{gaffmonk}, and a non-dimensional domain size, $L^2=4.5$. The latter allows larger numbers of stripes to form should the system pattern compared to the domain length scales used in \cite{gaffmonk}, enabling the prospect of greater sensitivity of the final pattern when comparing different distributions of time delay, since pattern mode selection is particularly sensitive on larger domains \cite{crampin}.  We primarily use no flux (homogeneous Neumann) boundary conditions on the boundary of the spatial domain, namely
\begin{equation}\label{neumannbc}
    \frac{\partial u}{\partial x}=\frac{\partial v}{\partial x}=0, \quad \quad x=0,1,
\end{equation}
though we will consider an alternative set of boundary conditions in Section \ref{BCs_ICs_Sect}. 

 The functions $f$ and $g$ are the reaction kinetics, which we will vary in subsequent sections. We will consider kinetics that have a unique positive homogeneous steady state, $(u_\star,v_\star)$. Unless otherwise stated, initial conditions $(u_0,v_0)$ are chosen as a small random Gaussian perturbation from this homogeneous steady state given by
\begin{equation}\label{firstic}
\begin{pmatrix}u_0\\v_0\end{pmatrix}=\begin{pmatrix}u_\star(1+r_u(x))\\v_\star(1+r_v(x))\end{pmatrix},
\end{equation}
where $r_u(x)$, $r_v(x)$ are Normally distributed random variables for each $x \in [0,1]$ with zero mean and standard deviation $\sigma_{IC}$. Unless otherwise mentioned, we take $\sigma_{IC}=0.01$ throughout.

The stochastic nature of gene expression delays leads us to consider a mean-field approach to modelling the time delay \cite{bratsun,krausenew}. The function $K$ is the kernel of our distributed delay, representing the distribution of delay times from the underlying process. The functions $F$ and $G$ model reaction steps which incorporate this delay. By assuming each individual mechanism within the gene expression process occurs independently and identically, we use the central limit theorem to model the delay as a (truncated) Gaussian distribution, $K(s;\bm{p})$, with $s$ the integral variable $s\in[\tau_1,\tau_2]$, $0<\tau_1<\tau_2$ and $\bm{p}$ the vector of distribution parameters. We will also consider non-Normal distributions by investigating skewed Gaussian distributions as well.

\subsection{Symmetric Truncated Gaussian Formulation}\label{dist_sect}
Using a symmetric truncated Gaussian kernel, we have $\bm{p}=(\tau,\sigma)$, for some mean delay $\tau$ and standard deviation $\sigma$, with integration limits chosen as $\tau_1=\tau-n\sigma$ and $\tau_2=\tau+n\sigma$ for some $n\in\mathbb{N}$, such that $\tau_1=\tau-n\sigma>0$ to ensure positive time delays only. The distribution is then given by  \begin{equation}
K(s;\bm{p})=K_N(s;\tau,\sigma)=\Phi_c\frac{1}{\sigma\sqrt{2\pi}}\exp\left(-\frac{1}{2}\left(\frac{s-\tau}{\sigma}\right)^2\right),
\end{equation}
where $\Phi_c$ denotes the truncation scaling constant, which ensures that $K_N(s;\tau,\sigma)$ integrates to $1$ over the given integration domain $[\tau_1,\tau_2]$, and is computed as
\begin{equation}
    \Phi_c=\frac{1}{\phi\left(\frac{\tau_2-\tau}{\sigma}\right)-\phi\left(\frac{\tau_1-\tau}{\sigma}\right)},
\end{equation}
with $\phi(x)$ the cdf of the standard Gaussian distribution,\footnote{The error function is given by $\text{erf}(x)=\frac{2}{\sqrt{\pi}}\int_0^xe^{-z^2}\ dz$.}
\begin{equation}\label{phi}
    \phi(x)=\frac{1}{2}\left(1+\text{erf}\left(\frac{x}{\sqrt{2}}\right)\right).
\end{equation}
 Throughout this paper, we use $n=3$ so that the integration limits are $\tau_1=\tau-3\sigma$ and $\tau_2=\tau+3\sigma$. This was chosen so that a relatively large $\sigma$ value could be used for each $\tau$ while maintaining $\tau_1>0$. For each $\tau$ value, a maximum $\sigma$ value can be computed such that $\tau_1=\tau-3\sigma\geq0$ as $\sigma_{\max}=\frac{\tau}{3}$. By setting $\sigma<\sigma_{\max}$, we ensure that the integration domain strictly considers positive time delays only.

As $\sigma \to 0$, we have that $K_N(s;\tau,\sigma)\to\delta(s-\tau)$.
Using the sifting property of the delta function, we remark that the continuous time delay problem formulation in \eqref{genform} reduces in the limit of $\sigma \to 0$ to a discrete fixed delay case with delay $\tau$, where
$$
\F(u,v)=F(u_\tau,v_\tau), \quad \quad \G(u,v)=G(u_\tau,v_\tau).
$$
 Hence, the continuously distributed time delay model is a generalisation of the fixed time delay model.

\subsection{Skewed Gaussian Formulation}
We will also study the role that the shape of a distribution has by considering a skewed Gaussian kernel, using the parameters $\bm{p}=(\mu,\omega,\rho)$. Such a distribution exhibits asymmetry but is otherwise a relatively simple generalization of the Normal distribution. The probability density function of the skewed truncated Gaussian distribution is given by
\begin{equation}
    K(s;\bm{p})=K_S(s;\mu,\omega,\rho)=\frac{\Psi_c}{\omega}\sqrt{\frac{2}{\pi}}\exp\left(-\frac{1}{2}\left(\frac{s-\mu}{\omega}\right)^2\right)\phi\left(\rho\frac{s-\mu}{\omega}\right),
\end{equation}
where $\phi(x)$ is the same as in \eqref{phi}. The parameter $\rho$ denotes the skew factor. The distribution is negatively skewed for $\rho<0$ and positively skewed for $\rho>0$. Finally, we have that $\Psi_c$ is the truncation scaling constant, given as
\begin{equation}
    \Psi_c=\frac{1}{\Psi\left(\frac{\tau_2-\mu}{\omega},\rho\right)-\Psi\left(\frac{\tau_1-\mu}{\omega},\rho\right)},
\end{equation}
with $\Psi(x,\rho)$ the cdf of a skewed Gaussian distribution, described by
\begin{equation}
    \Psi(x,\rho)=\phi(x)-2T(x,\rho).
\end{equation}
The function $T(x,\rho)$ denotes the Owen's T function \cite{owenst} and is written as an integral in the form
\begin{equation}
    T(x,\rho)=\frac{1}{2\pi}\int_0^\rho\frac{e^{-\frac{1}{2}x^2(1+s^2)}}{1+s^2}\ \text{d}s\quad -\infty<x,\rho<\infty.
\end{equation}
In the computational implementation of the skewed truncated Gaussian pdf, the integral $T(x,\rho)$ is resolved numerically using the composite Simpson's rule with $100,000$ discretisation points.

The parameters $\mu$ and $\omega$ no longer denote the mean and standard deviation of the distribution, but instead the location of the maximum and a scaling factor. To compare how the skewed distribution affects the onset of patterning compared to that of the fixed delay case, we consider the mean of the skewed distribution, $\tau$, which is given by
\begin{equation}\label{anmean}
    \tau=\int_{\tau_1}^{\tau_2}s K_S(s;\mu,\omega,\rho)\ \ \text{d}s.
\end{equation}
Following \cite{skewed}, the mean of the skewed truncated Gaussian distribution is computed as
\begin{equation}\label{computetau}
\tau=\mu+\omega\Psi_c\left[K_S(\tau_1;\mu,\omega,\rho)-K_S(\tau_2;\mu,\omega,\rho)+\frac{2\rho}{\hat{\rho}\sqrt{2\pi}}\left(\phi\left(\hat{\rho}\frac{\tau_2-\mu}{\omega}\right)-\phi\left(\hat{\rho}\frac{\tau_1-\mu}{\omega}\right)\right)\right],
\end{equation}
with $\hat{\rho}=\left(1+\rho^2\right)^{1/2}$.  We note that, for a given $\rho$, all of the terms on the right-hand side of \eqref{computetau}, namely $\omega$, $\Psi_c$,  and $K_S(s;\mu,\omega,\rho)$, can be written explicitly in terms of $\mu$. Equation \eqref{computetau} can therefore be solved implicitly for $\mu(\tau)$, for a given $\tau$.

\begin{figure}
    \centering
    \begin{subfigure}[t]{0.3\textwidth}
        \centering
        \includegraphics[width=1\textwidth]{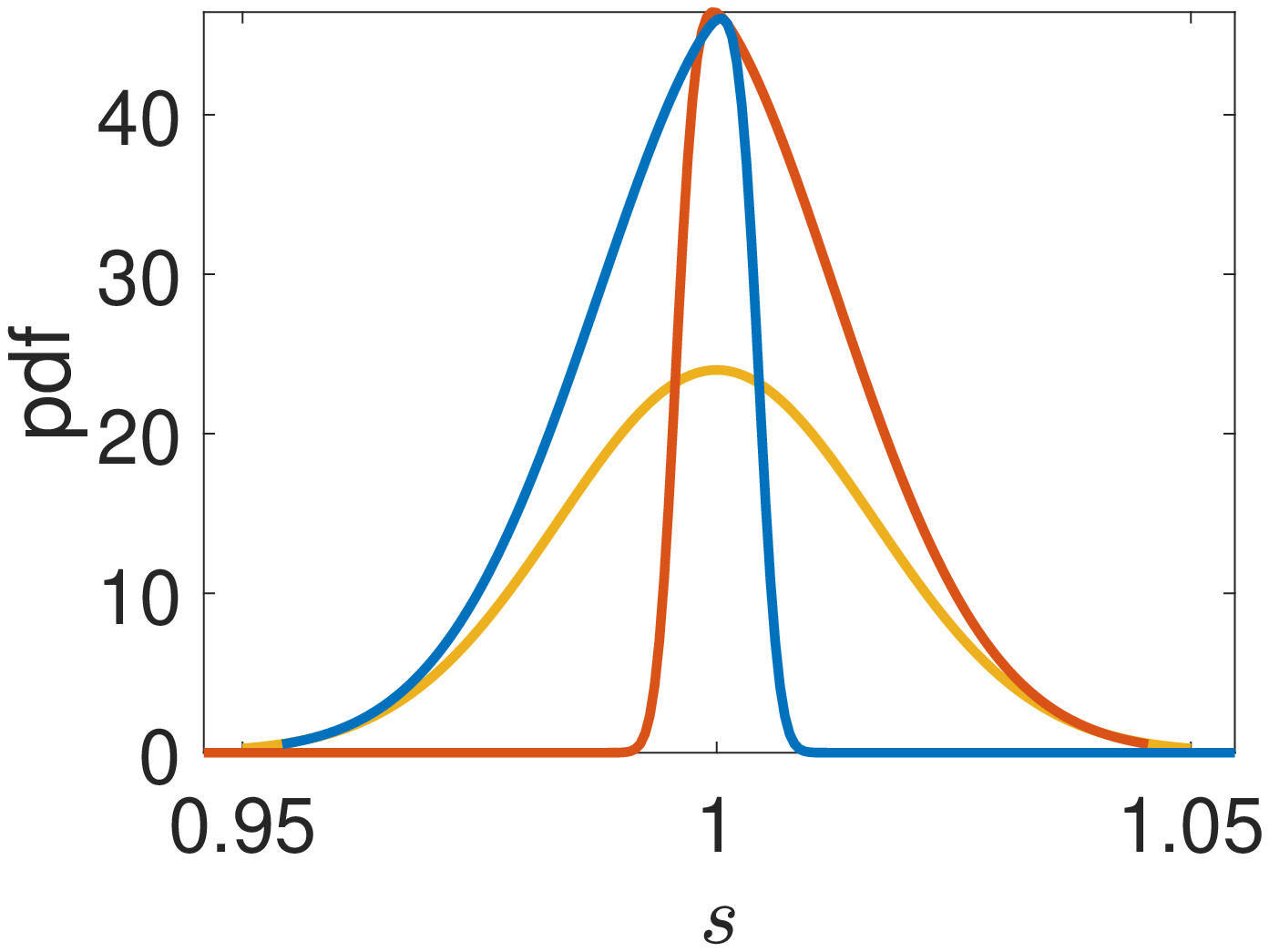}
        \caption{$\sigma=0.05\sigma_{\textrm{max}}$, $\omega=0.05\omega_{\textrm{max}}$}
    \end{subfigure}
        \begin{subfigure}[t]{0.3\textwidth}
        \centering
        \includegraphics[width=1\textwidth]{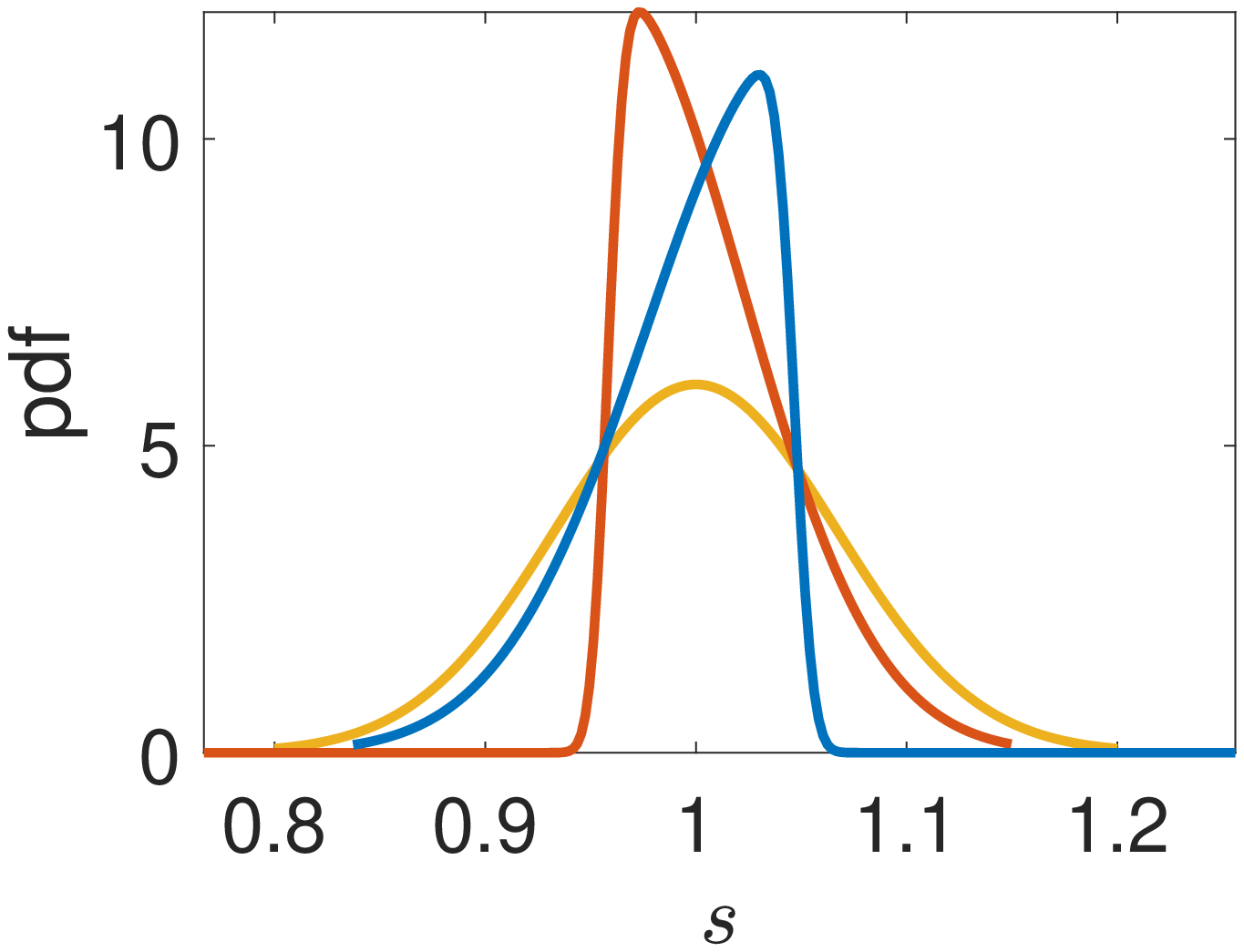}
    \caption{$\sigma=0.2\sigma_{\textrm{max}}$, $\omega=0.2\omega_{\textrm{max}}$}
    \end{subfigure}
        \begin{subfigure}[t]{0.3\textwidth}
        \centering
        \includegraphics[width=1\textwidth]{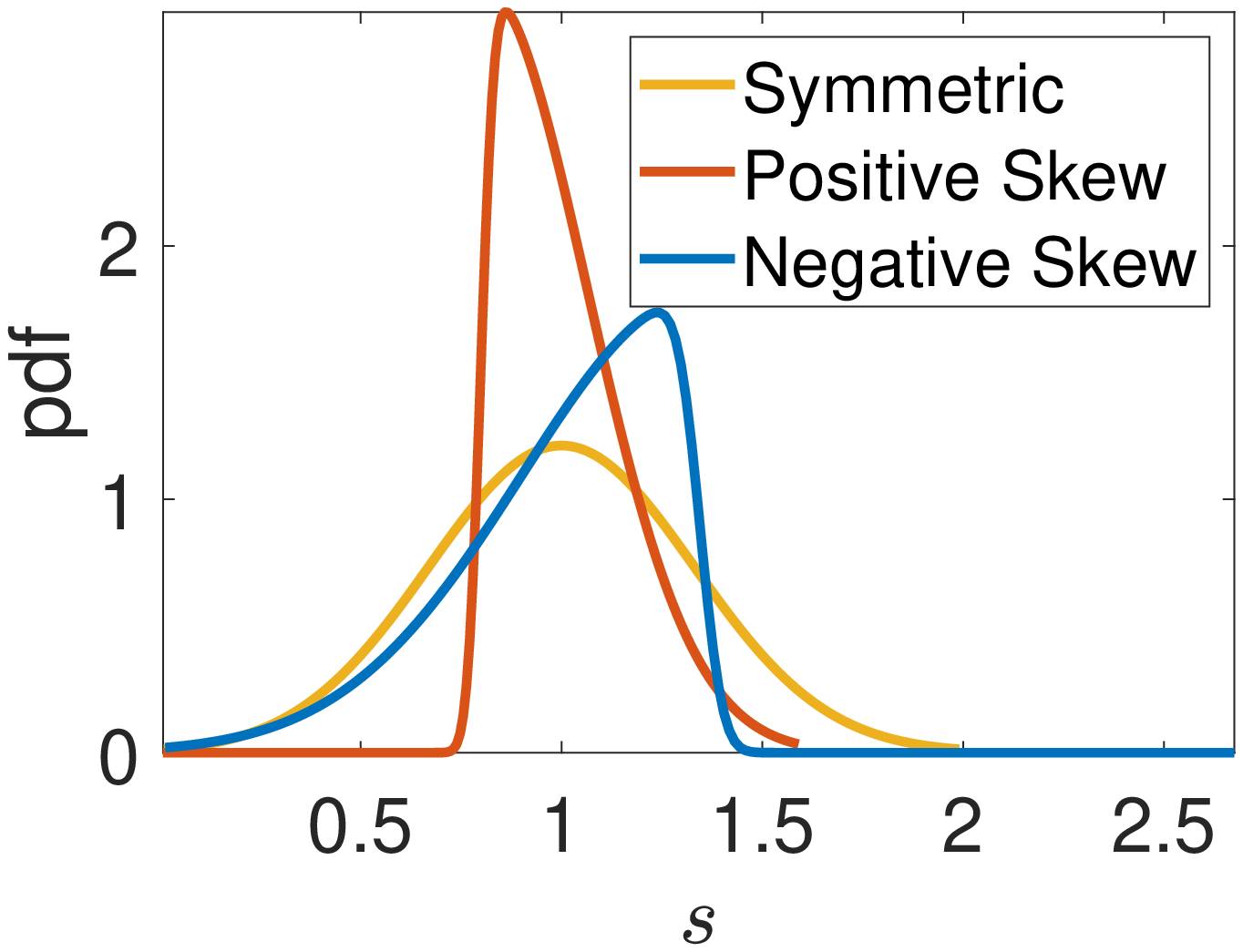}
    \caption{$\sigma=0.99\sigma_{\textrm{max}}$, $\omega=0.99\omega_{\textrm{max}}$}
    \end{subfigure}
    \caption{Probability density functions of the (symmetric) Normal, and Skew Normal distributions with a fixed mean of $\tau=1$ in all cases, and different standard deviations. The positively skewed distributions have $\rho=10$, and the negatively skewed $\rho=-10$. Note that the axes labels cover very different ranges, with small values of $\sigma$ and $\omega$ leading to very narrow distributions. For large values of $\omega$, the negative and positive skew cases are not symmetric about the mean due to the truncation used. For this value of $\tau$, we have that $\sigma_\textrm{max} \approx 0.3333$ and $\omega_\textrm{max} \approx 0.3354$. }
    \label{distribution_diagram}
\end{figure}

The integration limits were set to $\tau_1=\mu-3\omega$, $\tau_2=\mu+3\omega$, where $\omega$ was chosen such that $\omega<\omega_{\max}$, with $\omega_{\max}=\frac{\mu}{3}$ to ensure only positive time delays were considered. In Figure \ref{distribution_diagram} we give examples of these Skew Normal distributions compared to a standard symmetric Normal distribution, for different values of the standard deviation/scaling in each panel. 

\subsection{Reaction Kinetics}

We consider the Schnakenberg \cite{schnakenberg} and Gierer-Meinhardt (GM) \cite{gm} models as two canonical reaction-diffusion systems that exhibit Turing pattern formation. These models have also been studied extensively in the context of fixed time delays as described above. 

The stoichiometry of the Schnakenberg kinetics \cite{baker} is given by
\begin{equation}\label{chem}
A\xrightleftharpoons[c_{-1}]{c_1} U,\quad B\xrightarrow{c_2} V,\quad 2U+V\xrightarrow{c_3} 3U,
\end{equation}
where the $c_i$ represent reaction rates. The quantities $A$ and $B$ are reservoirs whose evolution is not considered, and we assume a constant supply rate. We use $u$, $v$, $a$, and $b$ to denote the (nondimensional) concentrations of substances $U$, $V$, $A$ and $B$ respectively, with $a$ and $b$ taken as constants. The form of the model we consider with time delay is the Ligand Internalistion (LI) variant of the standard Schnakenberg model. The LI model assumes that, a reaction at the cell surface is followed by internalisation of a morphogen, before the gene expression process can continue and morphogen production can occur \cite{leegaffney,yigaffneyli}. The time delayed terms in the LI model only appear in the activator's dynamics. This is based on the assumption that the gene expression process, and thus the source of the time delay, is responsible for autocatalysis of the activator in the reaction-diffusion mechanism \cite{gaffmonk}. Considering mass-action kinetics and adding standard diffusion terms \cite{murray}, we nondimensionalize the resulting system to find that the LI model with a distributed time delay is given by
\begin{equation}\label{distmodel2}
  \begin{split}
    \frac{\partial u}{\partial t}&=\frac{\epsilon^2}{L^2}\frac{\partial^2u}{\partial x^2}+a-u-2u^2v+3\int_{\tau_1}^{\tau_2}K(s;\textbf{p})u_s^2v_s \ \text{d}s,\\
    \frac{\partial v}{\partial t}&=\frac{1}{L^2}\frac{\partial^2v}{\partial x^2}+b-u^2v.
\end{split}
\end{equation}
The steady state of the Schnakenberg model is given by $(u_\star,v_\star)=\left(a+b,\frac{b}{(a+b)^2}\right)$. We do not consider the RLB variant proposed in \cite{leegaffney}, as it was shown to be able to exhibit negative concentrations for positive values of the feed rates $a$ and $b$ \cite{william}. We will also consider how this model compares to a fixed time delay LI model which is obtained in the limit $\sigma\to 0$.

Additionally, we consider delayed forms of the GM model in the fixed-delay case in order to explore the role of different kinetics and delay terms on our results. We will focus on two well-studied fixed time delay variants for simplicity, leaving an analysis of distributed delay in GM models to future work. A chemical interpretation of the kinetic reactions for the GM Model can be found in \cite{leegaffmonk}. The two non-dimensionalised model descriptions we consider, with kinetic reactions taken from \cite{murray}, and time delayed terms motivated by \cite{fadai1} and \cite{fadai2}, are given by \eqref{fadai1} and \eqref{fadai2}, and labelled GM$_1$ and GM$_2$ respectively, \footnote{We note that the papers \cite{fadai1,fadai2} label $v$ as the activator and $u$ as the inhibitor.}
\begin{multicols}{2}
\begin{equation}\label{fadai1}
  \left.\begin{split}
\frac{\partial u}{\partial t}&=\frac{\epsilon^2}{L^2}\frac{\partial^2 u}{\partial x^2}+a-bu+\frac{u_{\tau}^2}{v_{\tau}},\\
\frac{\partial v}{\partial t}&=\frac{1}{L^2}\frac{\partial^2 v}{\partial x^2}+u_{\tau}^2-v,
\end{split}\right\}\text{GM}_1
\end{equation}
\break
\begin{equation}\label{fadai2}
  \left.\begin{split}
\frac{\partial u}{\partial t}&=\frac{\epsilon^2}{L^2}\frac{\partial^2 u}{\partial x^2}+a-bu+\frac{u_{\tau}^2}{v},\\
\frac{\partial v}{\partial t}&=\frac{1}{L^2}\frac{\partial^2 v}{\partial x^2}+u_{\tau}^2-v,
\end{split}\right\}\text{GM}_2
\end{equation}
\end{multicols}
\noindent with $u_\tau=u(x,t-\tau)$ and $v_\tau=v(x,t-\tau)$. The key difference between these models is the inhibitor term in the activator's kinetics being delayed or not. The homogeneous steady state of the GM model is given by $$(u_\star,v_\star)=\left(\frac{a+1}{b},\left(\frac{a+1}{b}\right)^2\right).$$

\section{General Linear Instability Analysis}\label{Turing_Analysis_Sect}

We now consider the linear stability of homogeneous equilibria of the system \eqref{genform}. Denoting the steady state as $(u_\star,v_\star)$, we consider a small perturbation, $u(x,t)=u_\star+\delta\xi(x,t)$, $v(x,t)=v_\star+\delta\eta(x,t)$, where $|\delta|\ll1$. We have that
$$
f(u_\star,v_\star,\F(u_\star,v_\star))=g(u_\star,v_\star,\G(u_\star,v_\star))=0.
$$
Truncating at $O(\delta)$, we thus find that perturbations evolve according to
\begin{equation}\label{linearised}
    \begin{split}
        \pdiff{\xi}{t}&=\frac{\epsilon^2}{L^2}\pdiff{^2\xi}{x^2}+\xi\pdiff{f}{u}+\eta\pdiff{f}{v}+\pdiff{f}{\F}\int_{\tau_1}^{\tau_2}K(s;\bm{p})\left[\xi_s\pdiff{F}{u}+\eta_s\pdiff{F}{v}\right]\text{d}s, \\
        \pdiff{\eta}{t}&=\frac{1}{L^2}\pdiff{^2\eta}{x^2}+\xi\pdiff{g}{u}+\eta\pdiff{g}{v}+\pdiff{g}{\G}\int_{\tau_1}^{\tau_2}K(s;\bm{p})\left[\xi_s\pdiff{G}{u}+\eta_s\pdiff{G}{v}\right]\text{d}s,
    \end{split}
\end{equation}
where all partial derivatives are evaluated at the steady state, and $\xi_s=\xi(x,t-s)$, $\eta_s=\eta(x,t-s)$. Substituting into \eqref{linearised} an ansatz of the form $(\xi,\eta)^T=e^{\lambda_k t}\cos(k\pi x) (\xi_0,\eta_0)^T$ yields a homogeneous linear system for $(\xi_0,\eta_0)^T$ given by

\begin{equation}
\underbrace{\begin{pmatrix}\lambda_k+\frac{\epsilon^2}{L^2}(k\pi)^2-\pdiff{f}{u}-\pdiff{f}{\F}\pdiff{F}{u}E_k(\lambda_k)& -\pdiff{f}{v}-\pdiff{f}{F}\pdiff{F}{v}E_k(\lambda_k)\\
-\pdiff{g}{u}-\pdiff{g}{G}\pdiff{G}{u}E_k(\lambda_k)& \lambda_k+\frac{1}{L^2}(k\pi)^2-\pdiff{g}{v}-\pdiff{g}{\G}\pdiff{G}{v}E_k(\lambda_k) \end{pmatrix}}_{\bm{M}}\begin{pmatrix}\xi_0\\\eta_0\end{pmatrix}=\begin{pmatrix}0\\0\end{pmatrix},
\end{equation}
with $$E_k(\lambda_k)=\int_{\tau_1}^{\tau_2}K(s;\bm{p})e^{-\lambda_ks}\ \text{d}s.$$ 

Looking for non-trivial solutions of this system, we set $\det(\bm{M})=0$ to compute values of $\lambda_k$. This leads to a characteristic equation (or dispersion relation) of the form,
\begin{equation}\label{disp_rel}
    \D_k=\lambda_k^2 + \alpha_k \lambda_k + \beta_k + (\gamma_k \lambda_k + \delta_k)E_k(\lambda_k) + \chi_k E^2_k(\lambda_k)  = 0,
\end{equation}
with the coefficients given as
\begin{equation}
    \begin{split}
        \alpha_k&=\left(\frac{\epsilon^2}{L^2}+\frac{1}{L^2}\right)(k\pi)^2-\pdiff{f}{u}-\pdiff{g}{v},\\
        \beta_k&=\left(\frac{\epsilon^2}{L^2}(k\pi)^2-\pdiff{f}{u}\right)\left(\frac{1}{L^2}(k\pi)^2-\diff{g}{v}\right)-\pdiff{f}{v}\pdiff{g}{u},\\
        \gamma_k&=-\pdiff{g}{\G}\pdiff{\G}{v}-\pdiff{f}{\F}\pdiff{\F}{u},\\
        \delta_k&=-\pdiff{g}{\G}\pdiff{\G}{v}\left(\frac{\epsilon^2}{L^2}(k\pi)^2-\pdiff{f}{u}\right)-\pdiff{f}{\F}\pdiff{\F}{u}\left(\frac{1}{L^2}(k\pi)^2-\pdiff{g}{v}\right)-\pdiff{f}{v}\pdiff{g}{G}\pdiff{G}{u}-\pdiff{g}{u}\pdiff{f}{F}\pdiff{F}{v},\\
        \chi_k&=\pdiff{f}{\F}\pdiff{\F}{u}\pdiff{g}{\G}\pdiff{\G}{v}-\pdiff{f}{F}\pdiff{F}{v}\pdiff{g}{G}\pdiff{G}{u}.
    \end{split}
\end{equation}
The coefficients $\alpha_k$, $\beta_k$, $\gamma_k$, $\delta_k$, and $\chi_k$ depend only on the intrinsic model parameters, which are independent of the time delay distribution and presented for the LI and the two GM models, GM$_1$ and GM$_2$, in table \ref{tab:coeffs}.

Hence, one may note that the only term in the dispersion relation  that varies with the distribution is $E_k(\lambda_k)$, which can be computed for a skewed Gaussian distribution via:
\begin{equation}\label{ekskew}
E_k=\int_{\tau_1}^{\tau_2}K_S(s;\mu,\omega,\rho)e^{-\lambda_k s}\ \text{d}s=\frac{\Psi_c}{\omega\sqrt{2\pi}}\bigintsss_{\tau_1}^{\tau_2}\left(1+\text{erf}\left(\rho\frac{s-\mu}{\omega\sqrt{2}}\right)\right)\exp\left(-\frac{1}{2}\left(\frac{s-\mu}{\omega}\right)^2-\lambda_ks\right)\text{d}s.
\end{equation}
Setting $\mu=\tau$, $\omega=\sigma$, and $\rho=0$ to consider a symmetric distribution yields:
\begin{equation}\label{eksymm}
E_k=\int_{\tau_1}^{\tau_2}K_N(s;\tau,\sigma)e^{-\lambda_k s}\ \text{d}s=\frac{\Phi_c}{2}\left[\exp\left(\frac{\lambda_k(\lambda_k\sigma^2-2\tau)}{2}\right) \text{erf} \left(\frac{\lambda_k\sigma^2+s-\tau}{\sqrt{2}\sigma}\right)\right]\Bigg|_{\tau_1}^{\tau_2}.
\end{equation}
Finally, taking $\sigma\to0$ results in
\begin{equation}\label{ekfixed}
    E_k=e^{-\lambda_k\tau}, 
\end{equation}
as expected, corresponding to the fixed delay case with time delay $\tau$.

The $E_k$ term represents the main impact of delay on linear stability. Even in the relatively simple fixed delay case, we note that the \eqref{disp_rel} is not a quadratic equation for $\lambda_k$, but a transcendental one. The more complicated forms of $E_k$ in the distributed cases can be evaluated numerically, which is in general necessary to analyze these transcendental dispersion relations. We also remark that while \eqref{ekskew} seemingly contains many parameters, all of these can be rewritten in terms of $\tau$ and $\sigma$, and hence compared directly with the symmetric case given in \eqref{eksymm} (though the expressions for $\rho \neq 0$ provide no obvious insight, and again one must resort to numerical computation of $\lambda_k$).

The characteristic equation \eqref{disp_rel} can be used to determine the parameter sets $(a,b,\epsilon^2,L^2,\tau, \sigma)$ in which a Turing instability occurs (the `Turing space') and hence where we expect pattern formation. If there exists a $k\neq0$ for a given set of parameters such that $\max_k(\Re(\lambda_k))>0$, then we expect pattern formation. The largest value of $\Re(\lambda_k)$ also gives some indication of how quickly a perturbation grows away from the homogeneous steady state, and hence gives a heuristic estimate of the time taken for pattern formation to start.
\begin{table}[h]
\centering
\begin{tabular}{lcccc}
 & \multicolumn{1}{l}{} & \textbf{LI} & \textbf{GM$_1$} & \textbf{GM$_2$} \\ \cline{2-5}
 & \textbf{Parameter} & \textbf{} & \textbf{} & \textbf{} \\
 & $\alpha_k$ & $\left(\frac{\epsilon^2}{L^2}+\frac{1}{L^2}\right)k^2\pi^2+u_\star^2+4u_\star v_\star+1$ & $\left(\frac{\epsilon^2}{L^2}+\frac{1}{L^2}\right)k^2\pi^2+b+1$ & $\left(\frac{\epsilon^2}{L^2}+\frac{1}{L^2}\right)k^2\pi^2+b+1$ \\
 & $\beta_k$ & $\left(\frac{1}{L^2}\pi^2k^2+u_\star^2\right)\left(\frac{\epsilon^2}{L^2}\pi^2k^2+4u_\star v_\star+1\right)-4u_\star^3v_\star$ & $\left(\frac{\epsilon^2}{L^2}\pi^2k^2+b\right)\left(\frac{1}{L^2}\pi^2k^2+1\right)$ & $\left(\frac{\epsilon^2}{L^2}\pi^2k^2+b\right)\left(\frac{1}{L^2}\pi^2k^2+1\right)$ \\
 & $\gamma_k$ & $-6u_\star v_\star$ & $-2\frac{u_\star}{v_\star}$ & $-2\frac{u_\star}{v_\star}$ \\
 & $\delta_k$ & $-\frac{6}{L^2}u_\star v_\star k^2\pi^2$ & $-2\frac{u_\star}{v_\star}\left(\frac{1}{L^2}k^2\pi^2+1\right)$ & $-2\frac{u_\star}{v_\star}\left(\frac{1}{L^2}k^2\pi^2+1\right)+2\frac{u_\star^3}{v_\star^2}$ \\
 & $\chi_k$ & 0 & $2\frac{u_\star^3}{v_\star^2}$ & 0 \\ \cline{2-5}
\end{tabular}
\caption{A table of the coefficients of the characteristic equation \eqref{disp_rel} for the LI, GM$_1$, and GM$_2$ models. We note that these coefficients agree with those computed in \cite{yigaffneyli} for the LI model, up to a small typographical error in the coefficient of $\beta_k$.}
\label{tab:coeffs}
\end{table}

\section{Turing Spaces Under Delay}\label{Turing_Application_Sect}

In this Section we apply the results of the linear instability analysis to produce bifurcation diagrams of Turing pattern formation as the form of delay varies. Our results highlight the importance of the placement of time delay terms in the reaction kinetics, and show that this can alter the effect that delay has on the Turing space. We finally examine the effects that a continuously distributed delay has on the dominant eigenvalue of perturbation growth compared with that of a fixed delay with the same mean delay.

\subsection{Fixed Time Delay}
\begin{figure}
    \centering
    \begin{subfigure}[t]{0.45\textwidth}
        \centering
        \includegraphics[width=7cm,height = 6cm]{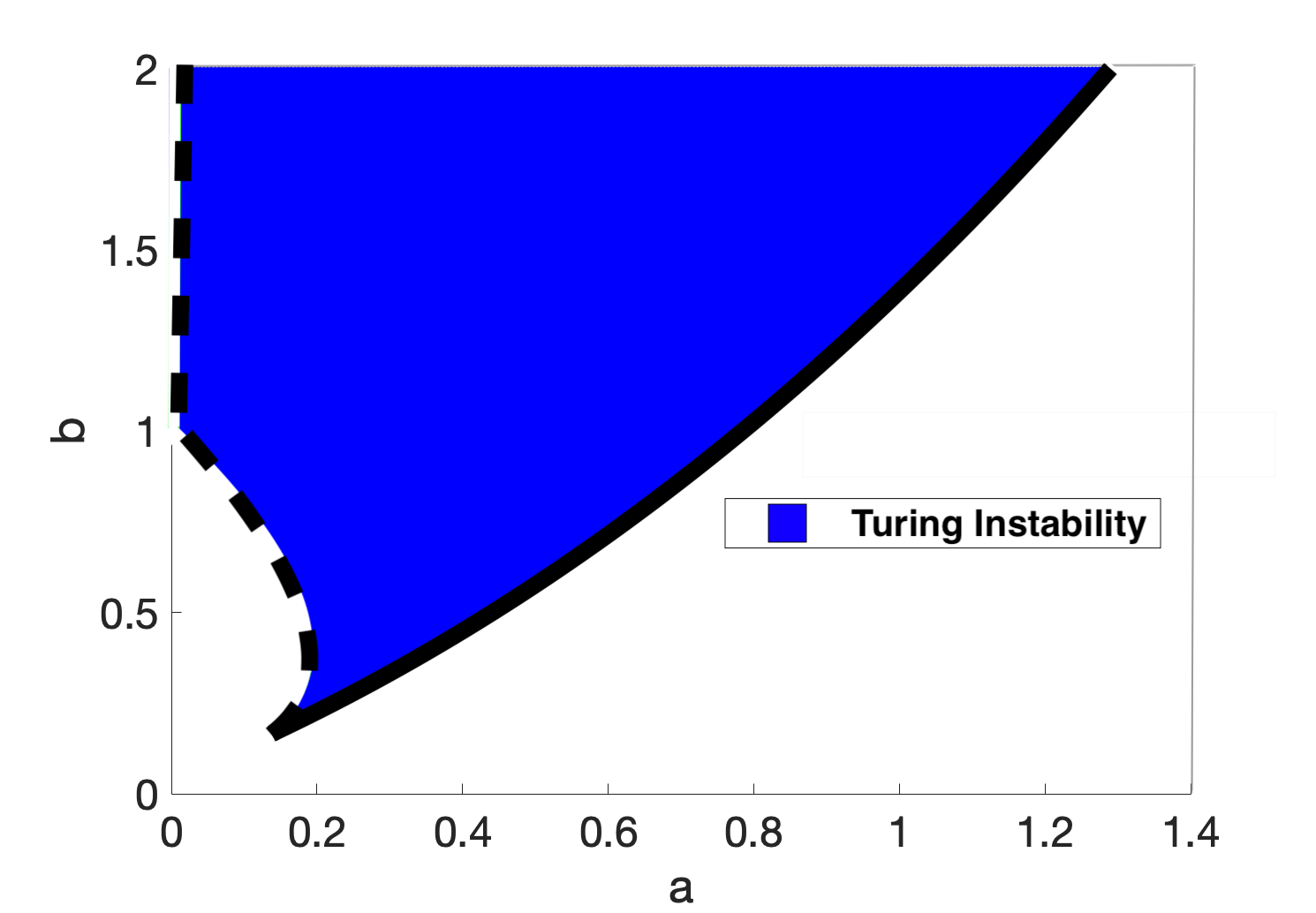}
        \caption{Turing space for Schnakenberg model.}
        \label{fig:tspace1}
    \end{subfigure}
    \hfill
    \begin{subfigure}[t]{0.45\textwidth}
        \centering
        \includegraphics[width=7cm,height = 6cm]{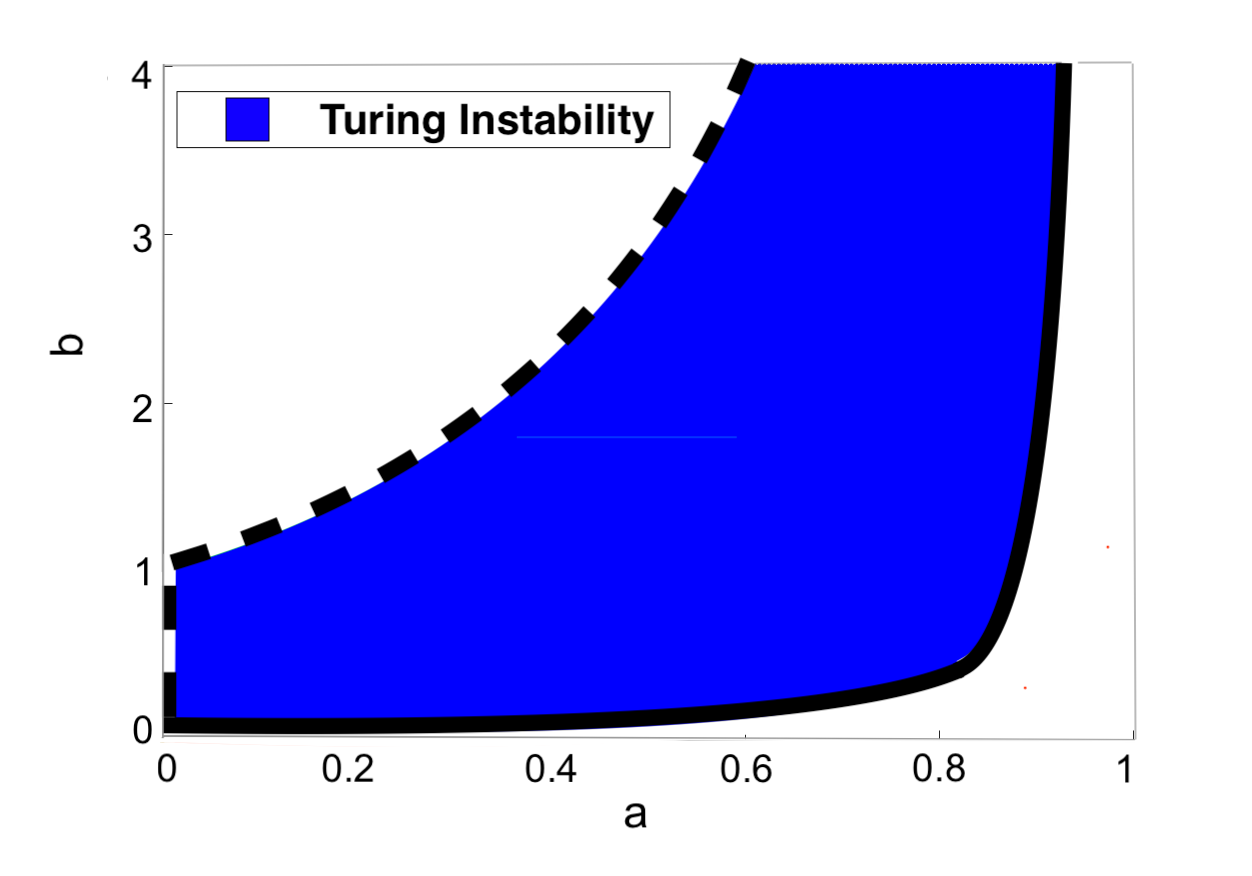}
        \caption{Turing space for GM model.}
        \label{fig:tspace2}
    \end{subfigure}
    \caption{Turing spaces for the Schnakenberg and GM models without delay, with $\epsilon^2=0.001$ and $L^2=4.5$. The dashed black arc (leftmost boundary) corresponds to parameters in which $\Re(\lambda_0)=0$, {\bl and the solid black outer arc (rightmost boundary) corresponds to parameters satisfying $\max_k(\Re(\lambda_k))=0$ for $k\neq0$.}}
    \label{fig:turingspaces}
\end{figure}

In Figure \ref{fig:turingspaces} we show Turing spaces of the Schnakenberg and GM models, which are bifurcation diagrams indicating regions of Turing instability in the absence of delay. We note two separate curves which separate the parameter space into its distinct regions. These will be referred to as the \textit{stability lines}, and are highlighted for the models in red and green. The inner green arc corresponds to the values of $(a,b)$ such that $\Re(\lambda_0)=0$ for the spatially homogeneous characteristic equation. {\bl The outer red boundary is comprised of the points $(a,b)$ such that $\max_k(\Re(\lambda_k))=0$, $k\neq0$.} We are interested in how these Turing spaces change as time delay is varied, and are also interested in the quantitative effects that changing delay has on the value of $\max_k(\Re(\lambda_k))$, as this can be thought of as a proxy for the time taken for patterns to form.

The roots of the characteristic equation were  found using the \emph{roots} command of the MATLAB package Chebfun \cite{chebfun}. In order to compute $\max_k(\Re(\lambda_k))$, we varied $k\in\mathbb{Z}$ over $[0,50]$ for a given $\tau$ and then took the maximum over $k$. We do not consider $k>50$ as full numerical solutions for the parameter values used tended towards patterns with four `spikes', so do not expect much larger wavenumbers to be excited (and full numerical simulations were used to check this assumption throughout, as well as to check the predictions of pattern formation).

\subsubsection{The LI Model}
\begin{figure}
    \centering
    \begin{subfigure}[t]{0.45\textwidth}
        \centering
        \includegraphics[width=7cm,height=5cm]{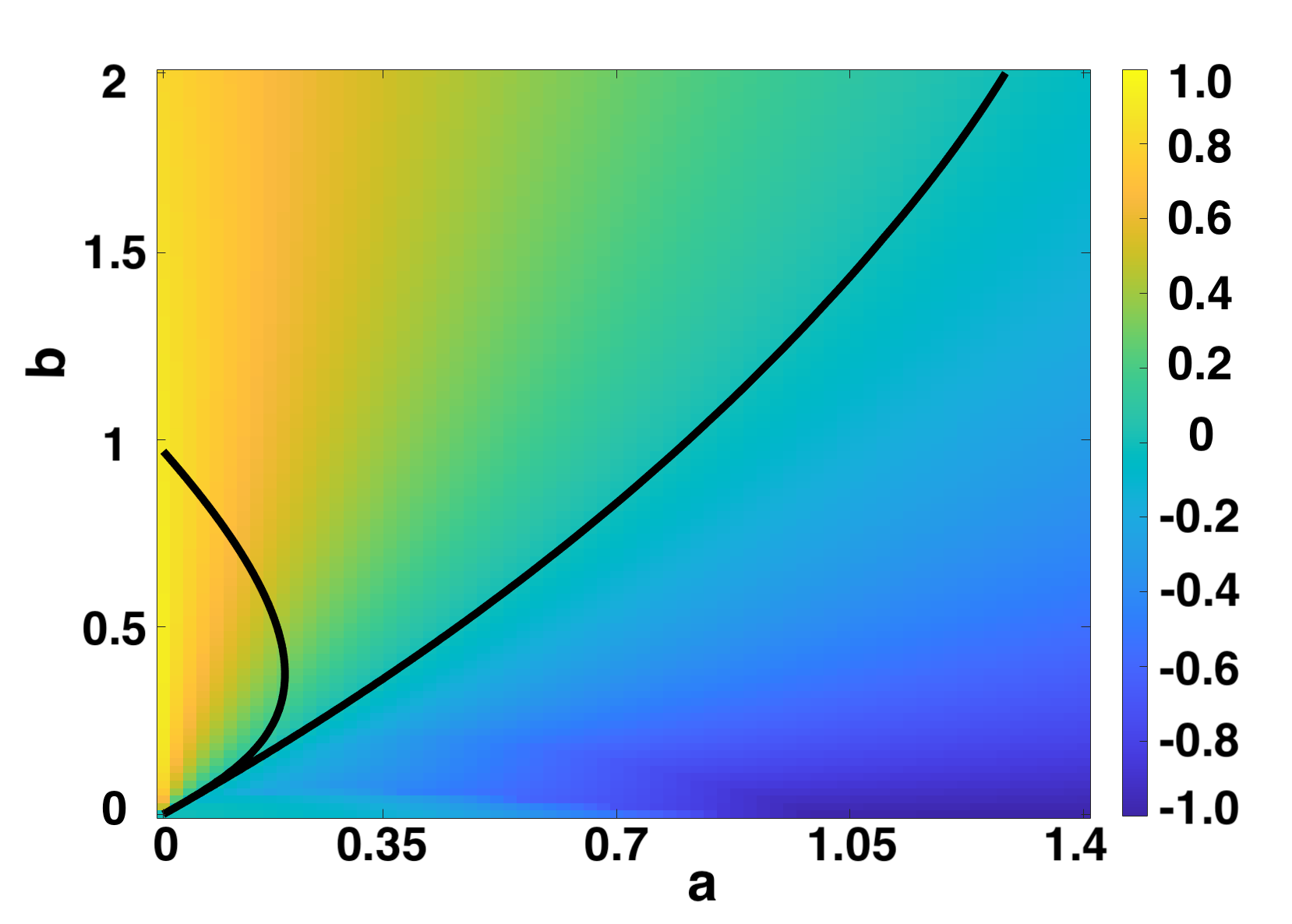}
        \caption{$\tau=0$.}
        
    \end{subfigure}
    \hfill
    \begin{subfigure}[t]{0.45\textwidth}
        \centering
        \includegraphics[width=7cm,height=5cm]{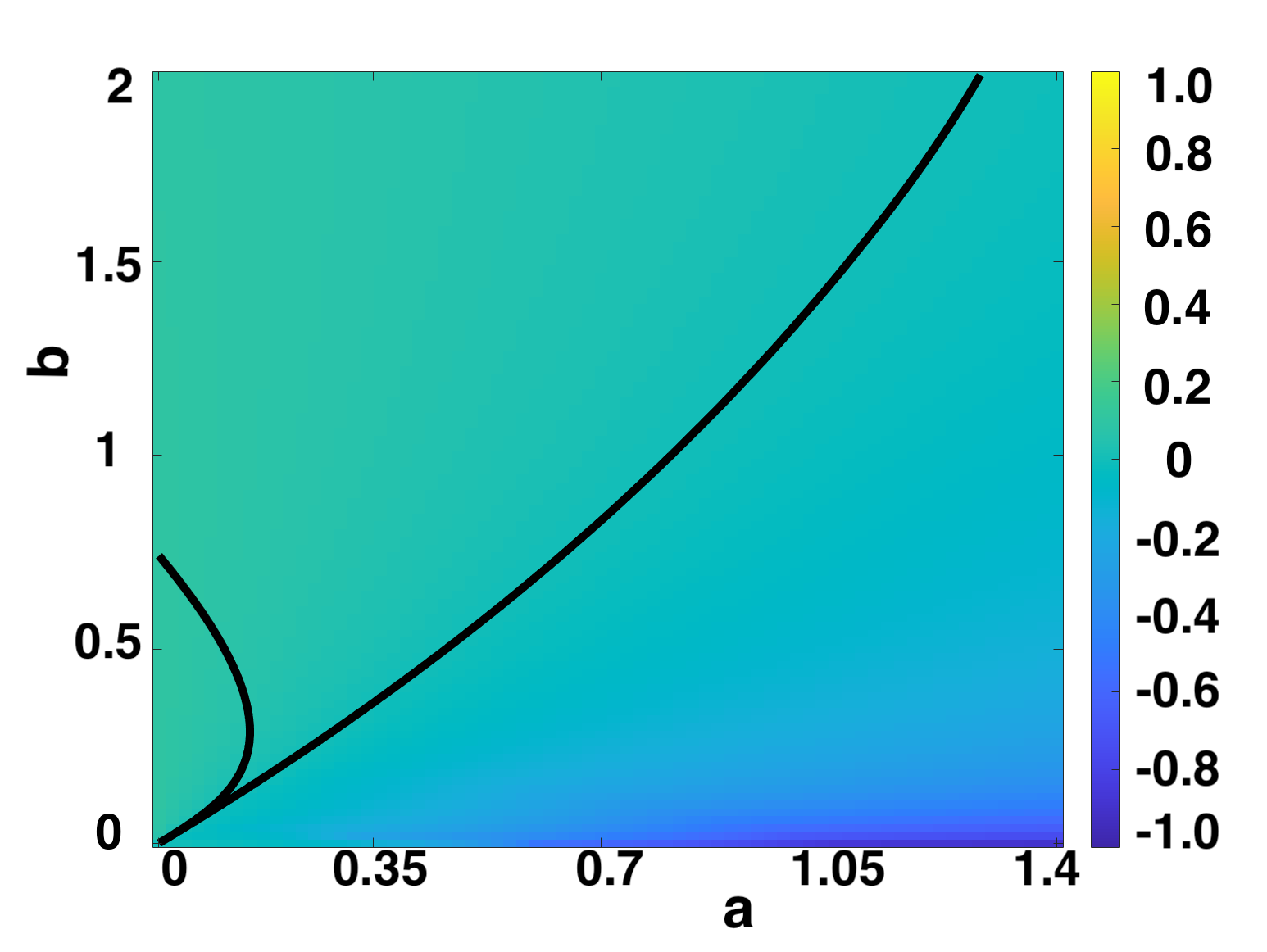}
        \caption{$\tau=1.5$.}
        
    \end{subfigure}
    \caption{$\max_k(\Re(\lambda_k))$ computed over $(a,b)$ parameter space by solving \eqref{disp_rel} for the fixed time delay LI model \eqref{distmodel2}, with $\epsilon^2=0.001$, $L^2=4.5$. As $\tau$ increases, $|\max_k(\Re(\lambda_k))|$ decreases. Stability lines for $\Re(\lambda_0)=0$ and $\max_k(\Re(\lambda_k))=0$, $k\neq0$, are overlaid, indicating the Turing space between them. }
    \label{fig:lambdavary}
\end{figure}
 Considering first the LI model with fixed time delay, in Figure \ref{fig:lambdavary} we plot a heatmap of $\max_k(\Re(\lambda_k))$ over the $(a,b)$ parameter space for the two fixed time delays $\tau=0$, $\tau=1.5$. We overlay contour lines corresponding to where $\Re(\lambda_0)=0$ and $\max_k(\Re(\lambda_k))=0$, $k\neq0$,
highlighting the Turing instability region. As $\tau$ increases, the region of homogeneous instability (leftmost curve in Figure \ref{fig:lambdavary} corresponding to $\Re(\lambda_0)=0$) decreases, increasing the Turing space. It can be seen that the absolute value $|\max_k(\Re(\lambda_k))|$ also decreases for $\tau = 1.5$. This suggests  pattern formation will take longer to occur, but this will happen over a larger Turing instability region of the $(a,b)$ parameter space. It similarly suggests that for $(a,b)$ such that $\max_k(\Re(\lambda_k))<0$, it will take a longer time for the eigenfunctions with modes $k\neq0$ to decay to a spatially homogeneous steady state. These results have been verified through full numerical solutions. %Figure \ref{fig:fixbif2} shows analogous bifurcation diagrams as in Figure \ref{fig:lambdavary}, but with $\epsilon^2=0.1$. We note that as the ratio of diffusion constants in the reaction-diffusion system, $\epsilon^2$, moves closer to $1$, the region of parameter space exhibiting Turing instability decreases. It can be observed however, that altering $\epsilon^2$ does not change the effect that an increasing $\tau$ has on $\max_k(\Re(\lambda_k))$, and that increasing the delay $\tau$ continues to act to promote Turing instabilities, with a shifting of the spatially homogeneous inner arc.
Furthermore the outer curve, corresponding to $\max_k(\Re(\lambda_k))=0$ for $k \neq 0$, does not  move at the resolution of the plotting  $\tau$ changes, while
analogous results (not shown) were found for other values of $\tau$ and $\epsilon$ \cite{sargood2022gene}.

\if 0  % Removes Figure 3, but if "false" with \fi the endif below 
\begin{figure}
    \centering
    \begin{subfigure}[t]{0.45\textwidth}
        \centering
        \includegraphics[width=7cm,height=5cm]{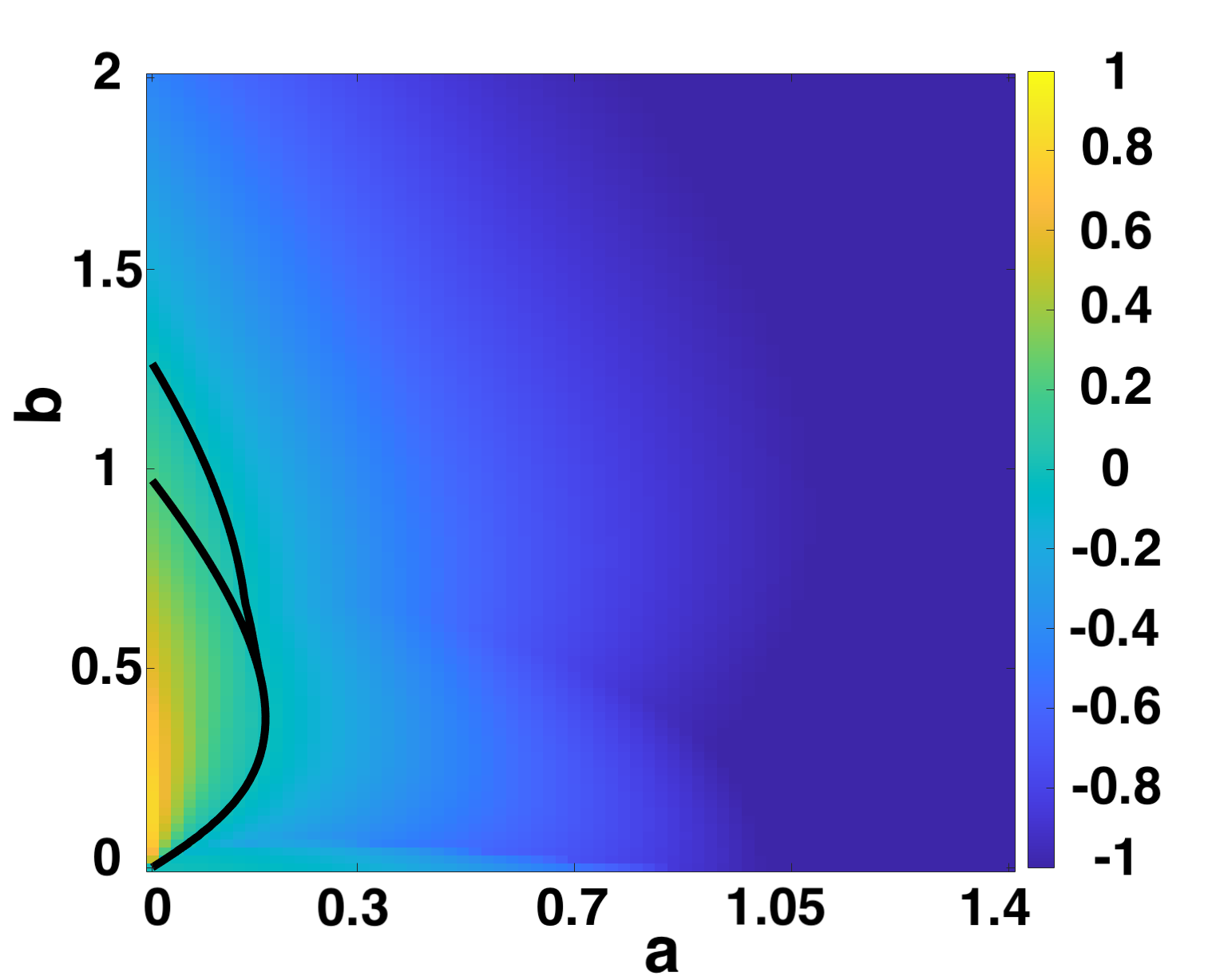}
        \caption{$\tau=0$.}
        
    \end{subfigure}
    \hfill
    \begin{subfigure}[t]{0.45\textwidth}
        \centering
        \includegraphics[width=7cm,height=5cm]{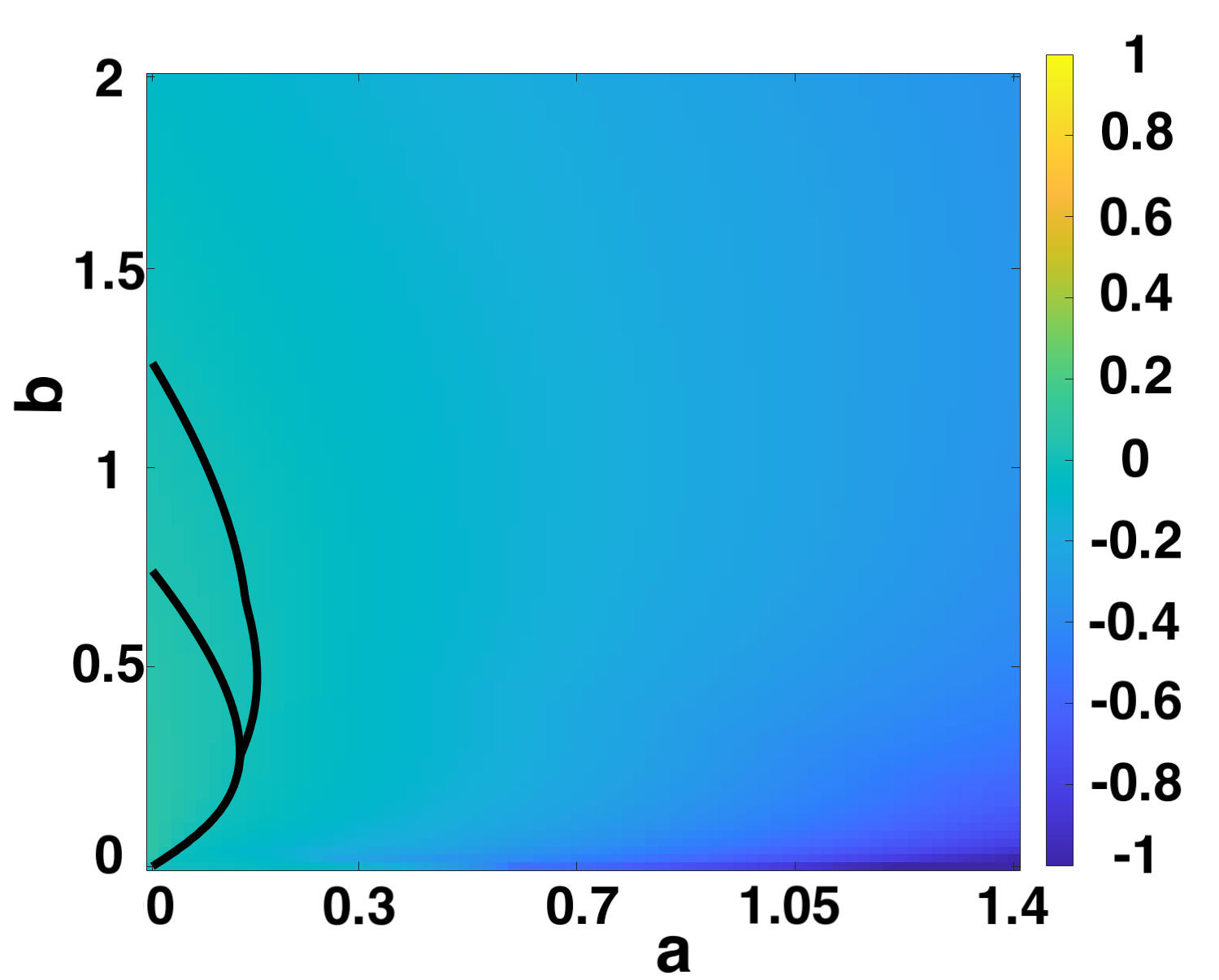}
        \caption{$\tau=1.5$.}
        
    \end{subfigure}
    \caption{$\max_k(\Re(\lambda_k))$ computed over $(a,b)$ parameter space by solving \eqref{disp_rel} for the fixed time delay LI model \eqref{distmodel2}, with $\epsilon^2=0.1$, $L^2=4.5$. As $\tau$ increases, $|\max_k(\Re(\lambda_k))|$ decreases. Stability lines for $\Re(\lambda_0)=0$ and $\max_k(\Re(\lambda_k))=0$ are overlaid, indicating the Turing space between them.}
    \label{fig:fixbif2}
\end{figure}
\fi 
These results were confirmed through full numerical solutions, and pattern formation was found where suggested by linear theory. We note that care ought to be taken when numerically simulating these models. The parameter region in the bottom left of the parameter space is a delicate region that can exhibit both Turing and (homogeneous) Hopf bifurcations, leading to complex spatio-temporal behaviours. This type of dynamics in reaction-diffusion systems has been studied more extensively in \cite{krausefixed,jiang} and many others.

\subsubsection{The GM Models}
The results in \cite{fadai1} showed that an increasing time delay in GM$_1$ had an antagonistic effect, shrinking the parameter space exhibiting stable spike solutions. In contrast, an increasing time delay for GM$_2$ caused an expansion of the stable spike solution parameter regime \cite{fadai2}. Here, we use our linear analysis of the spatially homogeneous steady states to examine how an increasing time delay will affect the Turing space for each of these variants. 

\begin{figure}
    \centering
    \begin{subfigure}[t]{0.32\textwidth}
        \centering
        \includegraphics[width=5.5cm,height = 5cm]{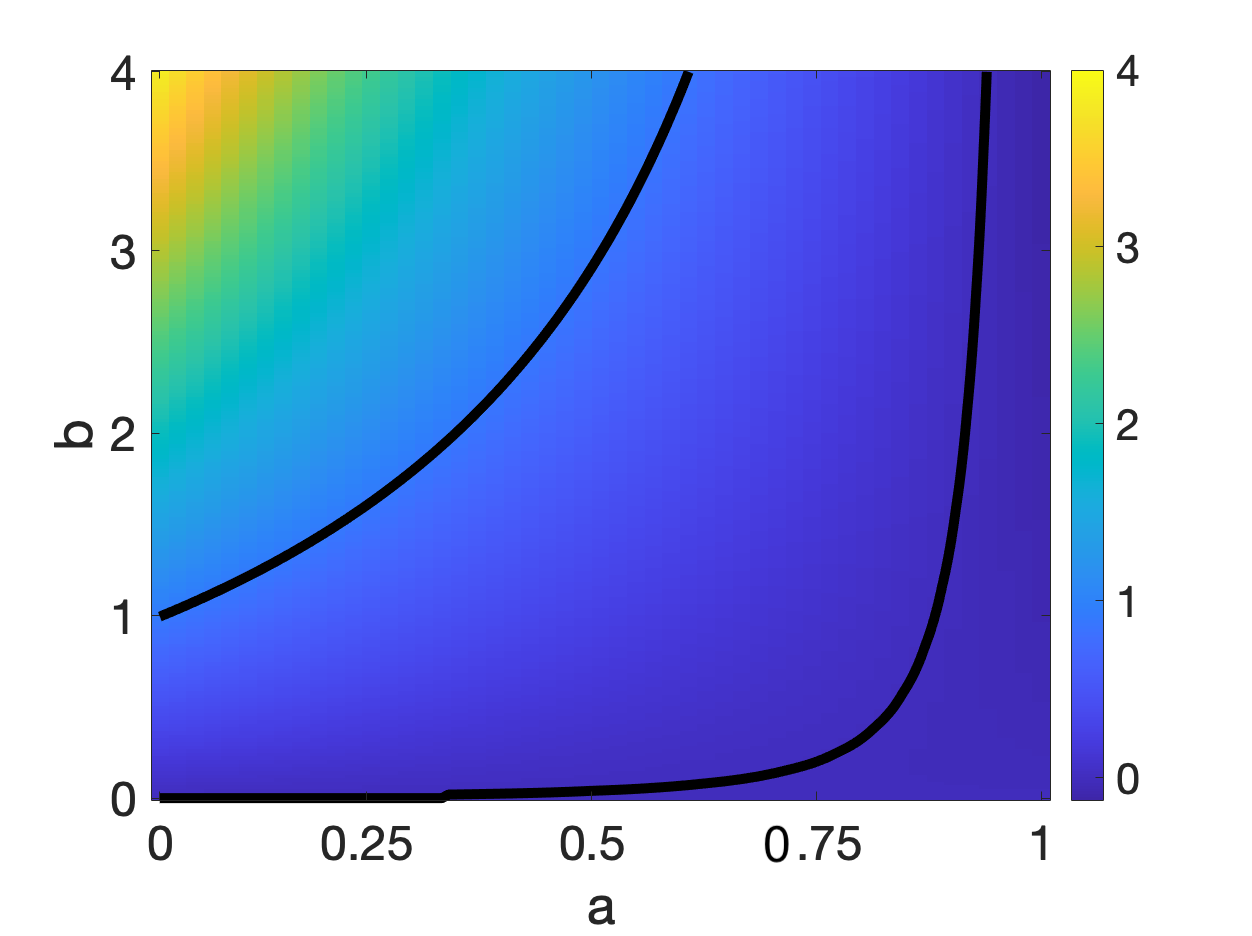}
        \caption{\bl{$\tau=0$} }
        
    \end{subfigure}
    \hfill
    \begin{subfigure}[t]{0.32\textwidth}
        \centering
        \includegraphics[width=5.5cm,height = 5cm]{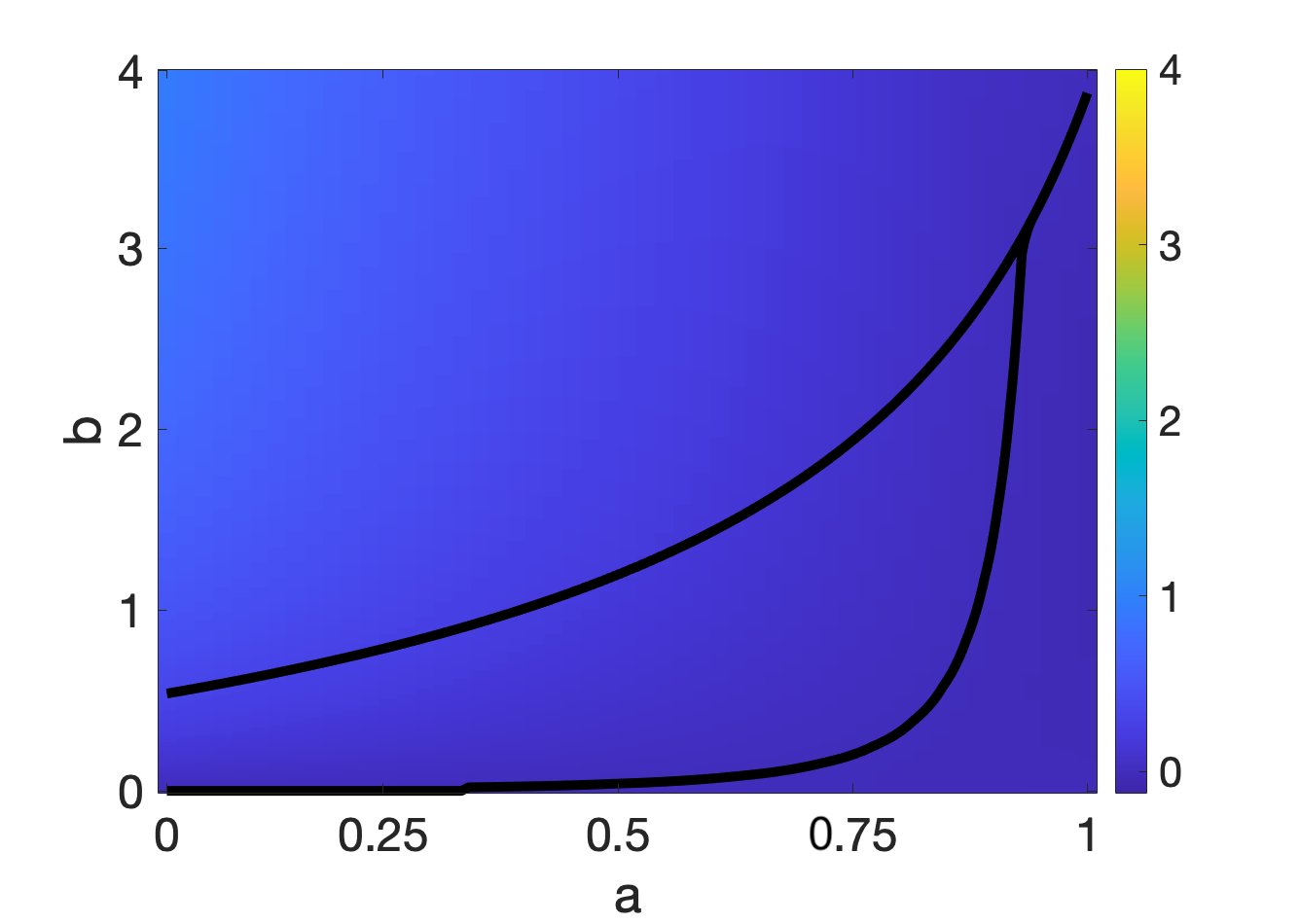}
        \caption{\bl{$\tau=0.5$ GM1 \eqref{fadai1}}}
        \label{fig:fad1}
    \end{subfigure}
    \hfill
    \begin{subfigure}[t]{0.32\textwidth}
        \centering
        \includegraphics[width=5.5cm,height = 5cm]{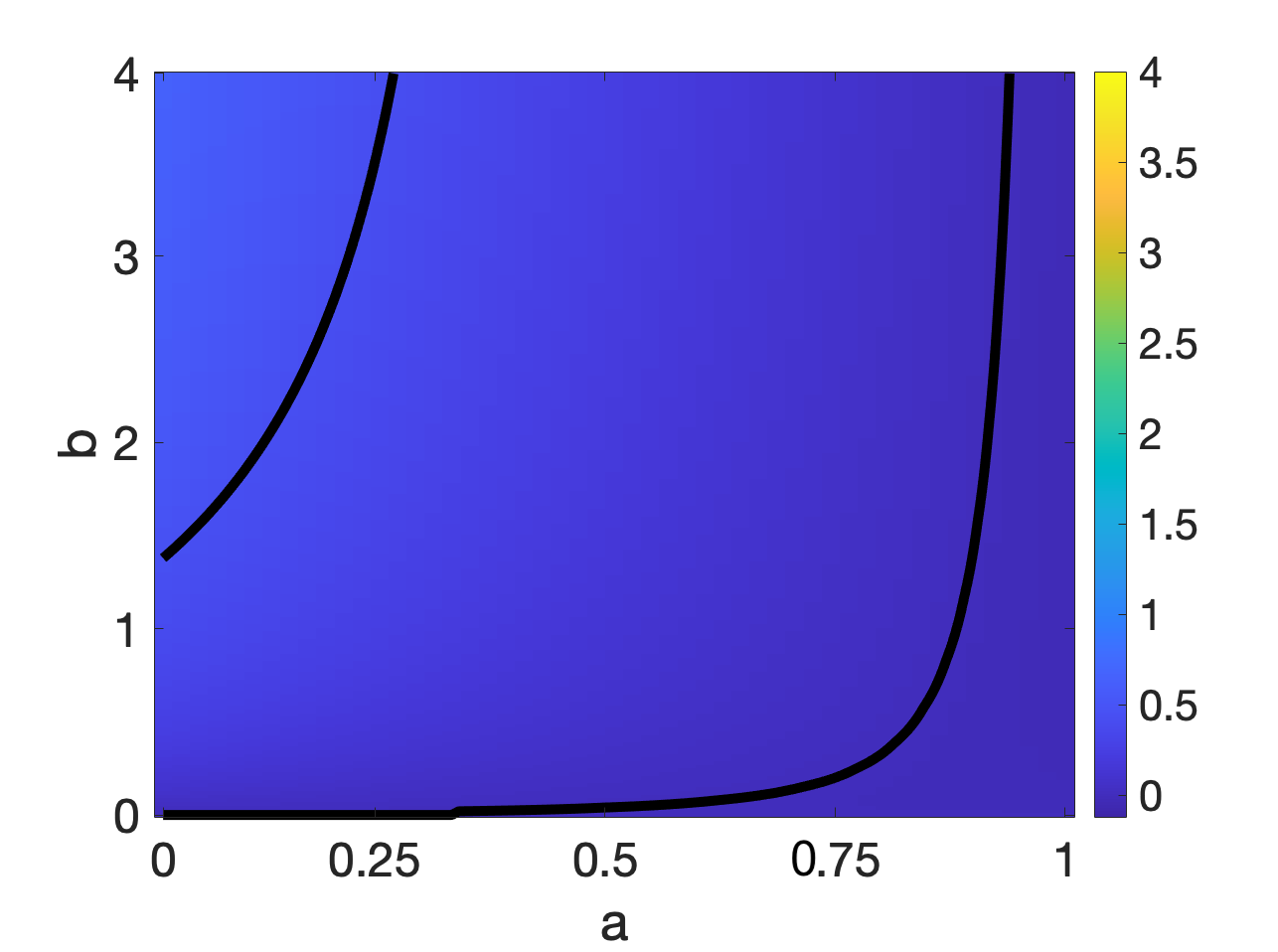}
        \caption{\bl{$\tau=0.8$ GM2 \eqref{fadai2}}}
        \label{fig:fad2}
    \end{subfigure}
    \caption{\bl{The maximum growth rate, $\max_k(\Re(\lambda_k))$, corresponding to solutions of \eqref{disp_rel} for models \eqref{fadai1} in (b) and \eqref{fadai2} in (c) with a fixed time delay, plotted for $(a,b)\in[0,1]\times[0,4]$, with varying $\tau$. Note that these models are equivalent for $\tau=0$ in (a). Parameters $\epsilon^2=0.001$ and $L^2=4.5$ used. Stability lines for $\Re(\lambda_0)=0$ and $\max_k(\Re(\lambda_k))=0$, $k\neq0$, are overlaid, indicating the Turing space between them.}}
    \label{fig:fadanalysis}
\end{figure}

Results in Figure \ref{fig:fad1} show a shrinking Turing space for increasing $\tau$ for GM$_1$, consistent with the analysis conducted in \cite{fadai1}, which showed a de-stabilisation of the stable spike solution parameter space with an increasing $\tau$. Similarly, results in \cite{fadai2} showed a stabilising effect of increasing $\tau$ on the stable spike solution parameter space for model GM$_2$ in \eqref{fadai2}, and we find an analogous result for the Turing space, shown in Figure \ref{fig:fad2}. In both cases we also observe a decrease in $|\max_k(\Re(\lambda_k))|$ as the delay is increased. We remark that the choice of delay times $\tau$ for these and other plots using the dispersion relation \eqref{disp_rel} was sufficiently large to demonstrate key trends for increasing $\tau$, but not so large that there were convergence issues in solving the transcendental dispersion relation (due to the multiple-scale nature of polynomial and exponential rootfinding). 

%\begin{figure}
 %   \centering
  %  \begin{subfigure}[t]{0.32\textwidth}
   %    \includegraphics[width=5.5cm,height = 5cm]{f1t0.png}
     %   \caption{$\tau=0$.}
    %    
    %\end{subfigure}
    %\hfill
    %\begin{subfigure}[t]{0.32\textwidth}
    %    \centering
     %   \includegraphics[width=5.5cm,height = 5cm]{f2t05.png}
     %   \caption{$\tau=0.5$.}
     %   
    %\end{subfigure}
    %\hfill
    %\begin{subfigure}[t]{0.32\textwidth}
    %    \centering
     %   \includegraphics[width=5.5cm,height = 5cm]{f2t08.png}
     %   \caption{$\tau=0.8$.}
     %   
    %\end{subfigure}
    %\caption{The maximum growth rate, $\max_k(\Re(\lambda_k))$, corresponding to solutions of \eqref{disp_rel} for \eqref{fadai2} with a fixed time delay, plotted for $(a,b)\in[0,1]\times[0,4]$, with varying $\tau$. Parameters $\epsilon^2=0.001$ and $L^2=4.5$ used. Stability lines for $\Re(\lambda_0)=0$ and $\max_k(\Re(\lambda_k))=0$ are overlaid, indicating the Turing space between them.}
    %\label{fig:fad2}
%\end{figure}

These results were also checked through full numerical solutions, and we found that the linear theory in all cases successfully predicted where Turing patterns would or would not form. These results indicate the importance of the positioning of time delayed terms within the kinetic reactions, as these differences in positioning of time delayed terms shows both expansion and contraction of the associated the Turing spaces.
    We finally note that in all the models considered, the ligand internalisation (LI) model and the two Gierer-Meinhardt (GM) models,  increasing $\tau$ changes the size of the Turing space only via  the  curve associated with the homogeneous characteristic equation ($k=0$), suggesting  a  specific mechanism by which gene expression time delays  impact  Turing space size.

\subsection{A Symmetric Gaussian Distributed Delay}
Next we consider the LI model with a symmetric Gaussian distribution for the time delay associated with the modelling of gene expression. The parameter $\sigma$ corresponds to a measure of the width of this distribution, with the limit $\sigma \to 0$ capturing the fixed time delay case. In particular $\sigma_{\max}$ has been defined to be the largest feasible value of $\sigma$ to avoid negative delays, as described in subsection \ref{dist_sect}. We consider the absolute difference of $\max_k(\Re(\lambda_k))$ for varying $\sigma$ values as a fraction of $\sigma_{\max}$, for multiple $\tau$ and $\epsilon$ values. For each $(\tau,\epsilon^2)$, bifurcation plots analogous to those of the previous subsection have been computed for the distributed delay case with varying $\sigma\in\{\sigma_{\max}\times0.99,\sigma_{\max}\times0.2,\sigma_{\max}\times0.1\}$.
However, the differences are below plotting resolution when compared to the fixed delay case, as may be found in \cite{sargood2022gene}; hence for each $(\tau,\epsilon^2)$ we instead  tabulate 
the absolute difference of $\max_k(\Re(\lambda_k))$ between each distributed delay case and the fixed delay case, across the $(a,b)$ parameter space as summarised in Table \ref{tab:tab1}.  

The largest absolute difference in $\max_k(\Re(\lambda_k))$ in Table \ref{tab:tab1} for all $\sigma$, $\tau$ and $\epsilon^2$ considered across the parameter space $(a,b)\in[0,1.4]\times[0,2]$ is $O(10^{-3})$. We therefore expect that for all $(a,b)\in[0,1.4]\times[0,2]$,
using a symmetric Gaussian distribution centred at some mean $\tau$ (for small $\tau$) will not significantly affect Turing instabilities compared to the fixed delay case, independent of the standard deviation $\sigma$ of the distribution. These results are a fairly robust indication that, at the level of linear instability, the distributed delay does not appreciably change the impact of time delay on Turing instability analysis.

\begin{table}
\centering
\begin{tabular}{lrrrr}
\hline
\multicolumn{2}{c}{Parameters Used}    & $\sigma_{\max}\times0.99$ & $\sigma_{\max}\times0.2\ $ & $\sigma_{\max}\times0.1\ $ \\ \hline
$\epsilon^2=0.001$ & \textbf{$\tau=0.2$} & $0.0010$                           & $4.2\times10^{-5}$                & $1.1\times10^{-5}$                \\
$\epsilon^2=0.001$ & $\tau=1.0$          & $0.0078$                           & $3.3\times10^{-4}$                & $8.2\times10^{-5}$                \\
$\epsilon^2=0.01$  & \textbf{$\tau=0.2$} & $0.0025$                           & $9.4\times10^{-5}$                & $2.3\times10^{-5}$                \\
$\epsilon^2=0.01$  & \textbf{$\tau=0.5$} & \textbf{$0.0076$}                  & $2.6\times10^{-4}$                & $6.4\times10^{-5}$               \\ \hline
\end{tabular}
\caption{Table showing the maximum value over the parameter spaces $(a,b)$ of the absolute difference of $\max_k(\Re(\lambda_k))$ between distributed delay cases and the fixed delay case, across the $(a,b)\in[0,1.4]\times[0,2]$ parameter space, for multiple $\tau$ and $\epsilon^2$ values. Throughout we set $L^2=4.5$. Results are displayed to $2$ significant figures.}
\label{tab:tab1}
\end{table}

\subsection{A Skewed Gaussian Distributed Delay}

\begin{figure}
    \centering
    \begin{subfigure}[t]{0.45\textwidth}
        \centering
        \includegraphics[width=7cm,height=5cm]{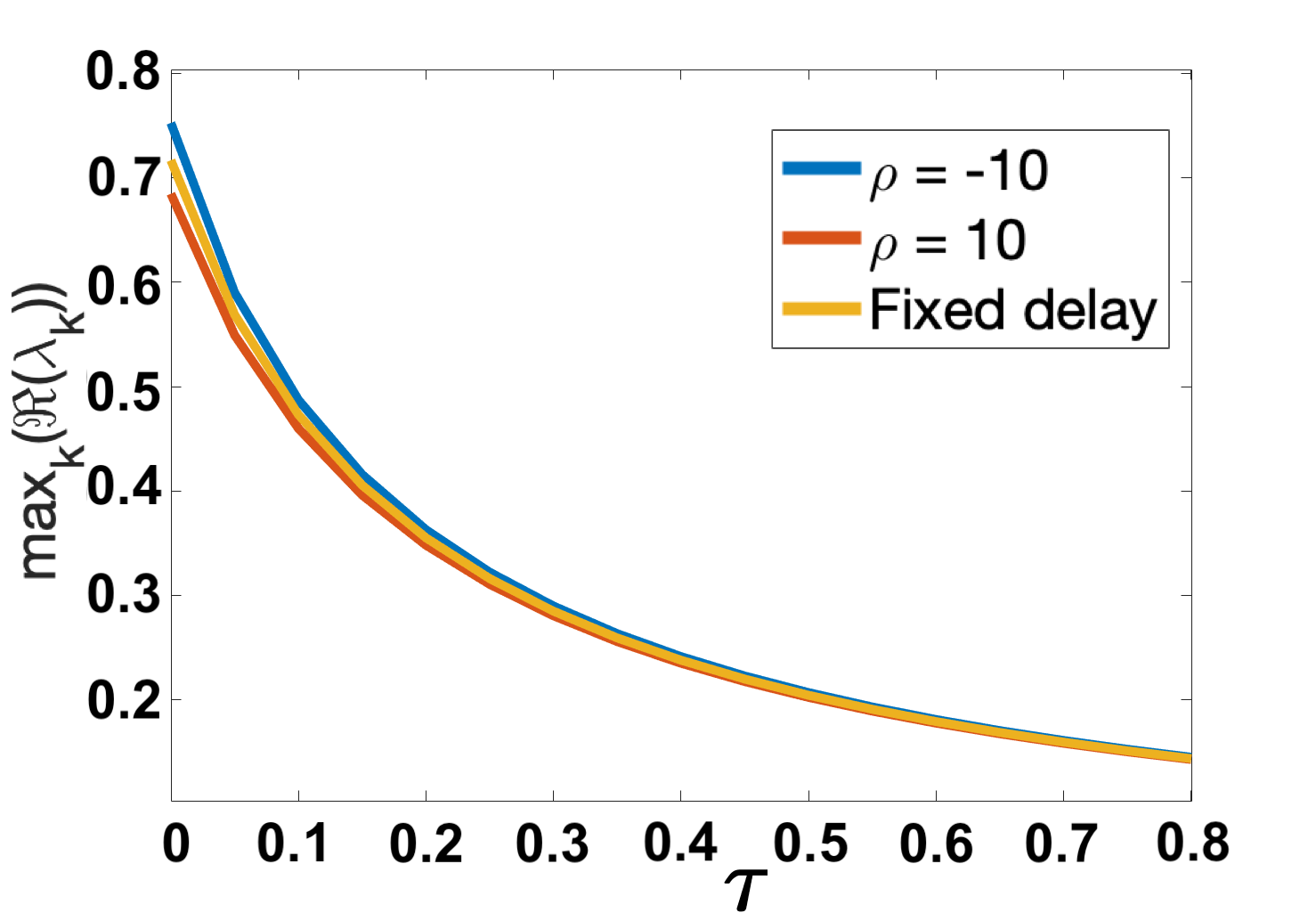}
        \caption{$(a,b)=(0.1,0.9)$.}
        
    \end{subfigure}
    \hfill
    \begin{subfigure}[t]{0.45\textwidth}
        \centering
        \includegraphics[width=7cm,height=5cm]{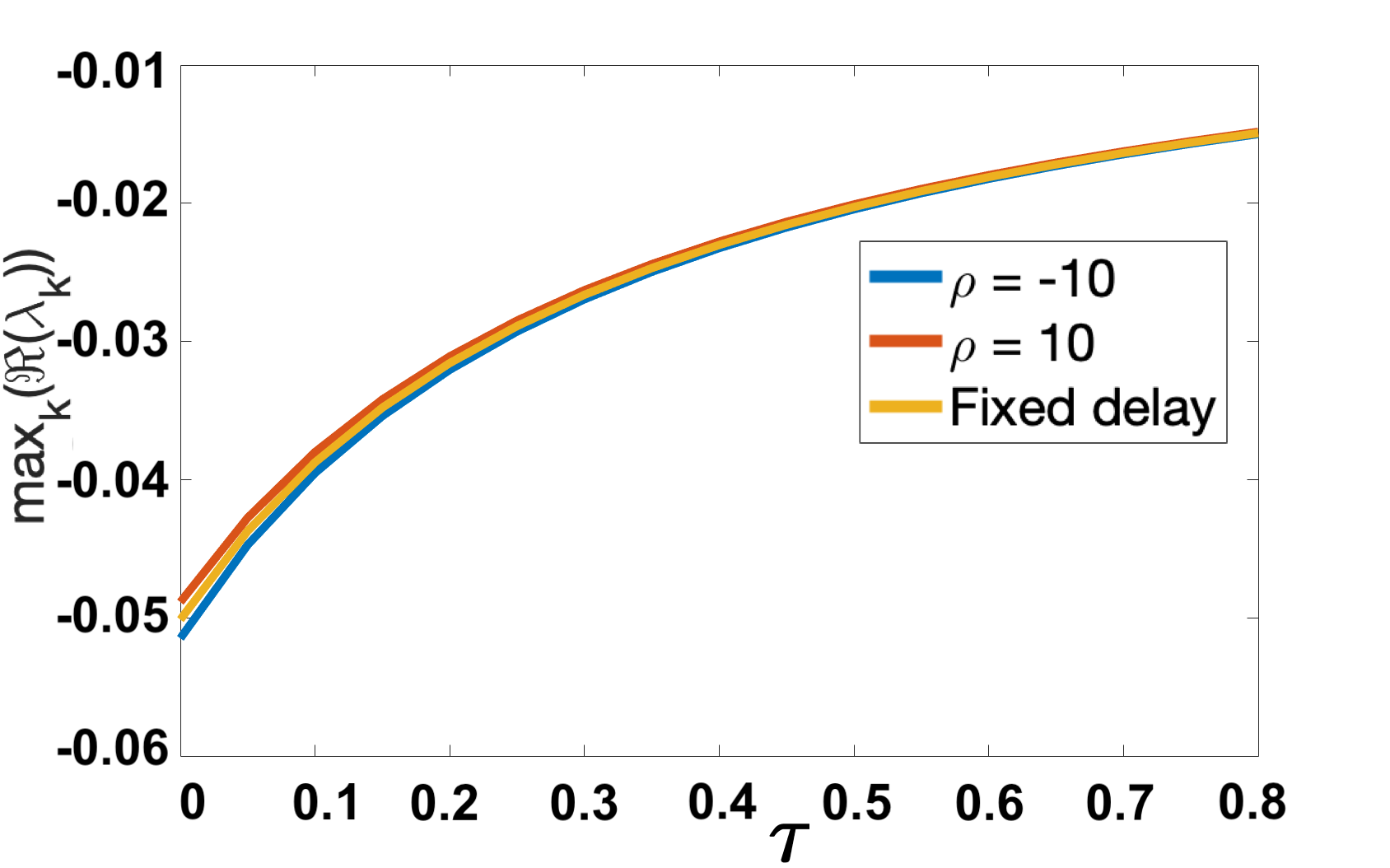}
        \caption{$(a,b)=(0.4,0.4)$.}
        
    \end{subfigure}
    \caption{Comparison of $\max_k(\Re(\lambda_k))$ computed from \eqref{disp_rel} plotted against $\tau\in[0,0.8]$ for $\rho=-10,10$ against the fixed delay case, with the distribution spread given by $\omega = \omega_{\max}\times0.99$. Parameter values $\epsilon^2=0.001$ and $L^2=4.5$ used. In generating the plot, the time delay scale, $\tau$, was varied at regular intervals of $0.05$ with $k\in\mathbb{Z}$ ranging over $k\in[0,50]$ before a maximum was taken.}
    \label{fig:dispskew}
\end{figure}

    Evaluating the dispersion relation \eqref{disp_rel} in the skewed distribution case is substantially more costly computationally, especially for larger values of $\tau$ where root-finding becomes numerically difficult. Instead, we plot dispersion relations for certain fixed $(a,b)$ parameter values close to the Turing space boundaries of Figure  \ref{fig:turingspaces}(a), and compare these plots to the fixed delay case. We show that for a small mean $\tau$, the skew, positive or negative, does not significantly affect the value of $\max_k(\Re(\lambda_k))$. We highlight here again that equation \eqref{computetau} can be solved implicitly for $\mu(\tau)$, for a given $\tau$, using the \textit{fzero} command in MATLAB. For a given $\rho$, and each found $\mu$, we compute $\max_k(\Re(\lambda_k))$ by solving for roots of the characteristic equation \eqref{disp_rel}. In Figure \ref{fig:dispskew}, we plot $\max_k(\Re(\lambda_k))$ against $\tau\in[0,0.8]$ for skew parameter values of $\rho=-10,10$, and $\omega=\omega_{\max}\times0.99$, with two different $(a,b)$ parameter sets. The value $\omega_{\max}$ is defined analogously to that of $\sigma_{\max}$. A plot of $\max_k(\Re(\lambda_k))$ for the fixed delay case is also added for comparison in each case. 

From Figure \ref{fig:dispskew} we see that the curves differ slightly for small $\tau$, with $\rho=-10$ having a slightly higher value of the maximum growth rate, and $\rho=10$ a slightly lower value. The overall effect is very small despite the large skew implemented in the distribution. Noting the overall unit scale of $|\max_k(\Re(\lambda_k))|$ in Figs.~\ref{fig:lambdavary},~\ref{fig:fad1},~\ref{fig:fad2}, we anticipate that these effects are small enough not to have a significant impact on the timescale for patterning onset, as confirmed numerically via simulations of the full model with distributed delay near the boundary of these instabilities.

\section{Numerical Exploration of the Models}\label{Numerical_Sect}

To verify our linear instability results, as well as explore the dynamics of pattern formation beyond the linear regime, full numerical solutions were conducted. The spatial derivatives were discretized via the method-of-lines using the standard three-point stencil for the Laplacian, with $m=500$ equally spaced points on the domain $x\in\Omega=[0,1]$. This discretization results in $m=500$ delay differential equations (DDEs) in time, which are solved via built-in time-stepping solvers in Julia, typically \emph{Rodas5}, a 5-th order A-stable solver, from the family of Rosenbrock methods \cite{rodas,rosenbrock}.  The distributed delay terms were approximated via a composite Simpson's rule using $50$ quadrature points; see \cite{sargood2022gene} for a more thorough discussion of the numerical procedures used. The Julia code used to generate all numerical solutions throughout this paper can be found at the open source repository \cite{git}.

With a maximum delay of $\tau_2=\tau+3\sigma$ (or $\mu+3\omega$ with a skewed-distribution), we note that to solve DDEs, a history function is required to define the solution for $t \in [-\tau,0)$ for the fixed delay case, $t\in[-\tau-3\sigma,0)$ for the symmetric distributed case, and $t\in[-\mu-3\omega,0)$ for the skewed distributed case. Unless otherwise stated, a constant history function equal to the initial conditions is used.
Finally, as the LI model has \textit{cross} reaction kinetics, we have that when the concentration of the activator $u$ is high, the concentration of the inhibitor $v$ is low, and vice-versa \cite{murray}. The concentration gradients of the two morphogens $u$ and $v$ are thus effectively `out of phase', and so it is sufficient to show just the numerical solution of the activator $u$. Hence, only the numerical solution of the activator $u$ is plotted.

\subsection{A Relationship between Fixed Delay and Onset of Patterning}

Here we show that, for small $\tau$ and $L^2$, the linear theory provides a good approximation to the time taken until pattern formation occurs, and in fact, the relationship between $\tau$ and time-to-pattern under these conditions is linear for the fixed delay case. We also show that, through full numerical solutions, the relationship between $\tau$ and time-to-pattern on a longer timescale for larger $\tau$ is also linear. We first consider the former.

\begin{figure}
    \centering
    \begin{subfigure}[t]{0.45\textwidth}
        \centering
        \includegraphics[width=7cm,height=5cm]{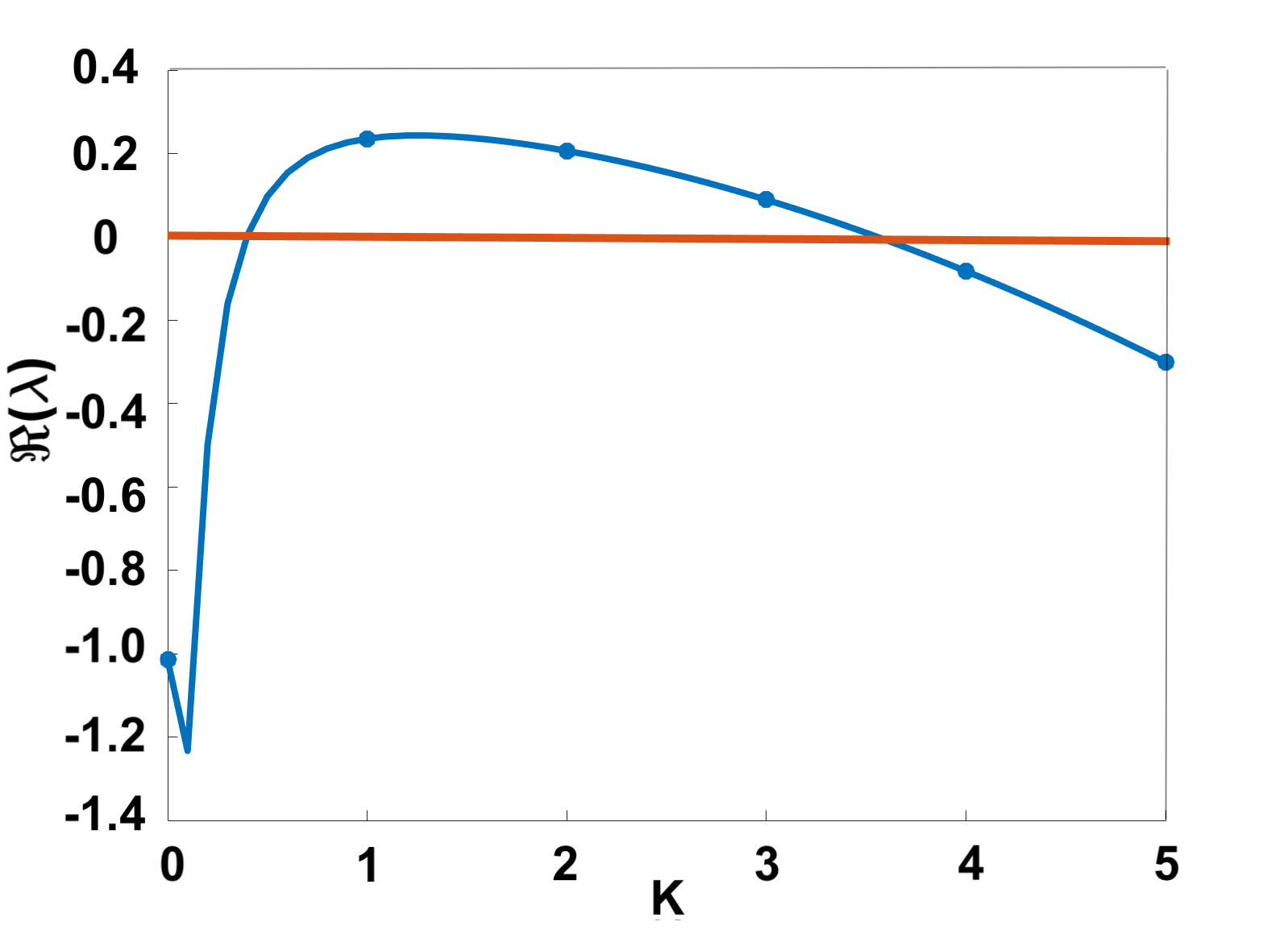}
        \caption{Dispersion curve plotted with domain size $L^2=0.2$. Curve produced by varying $k\in[0,5]$ at regular intervals of $1$.  }
        \label{fig:compdisp1}
    \end{subfigure}
    \hfill
    \begin{subfigure}[t]{0.45\textwidth}
        \centering
        \includegraphics[width=7cm,height=5cm]{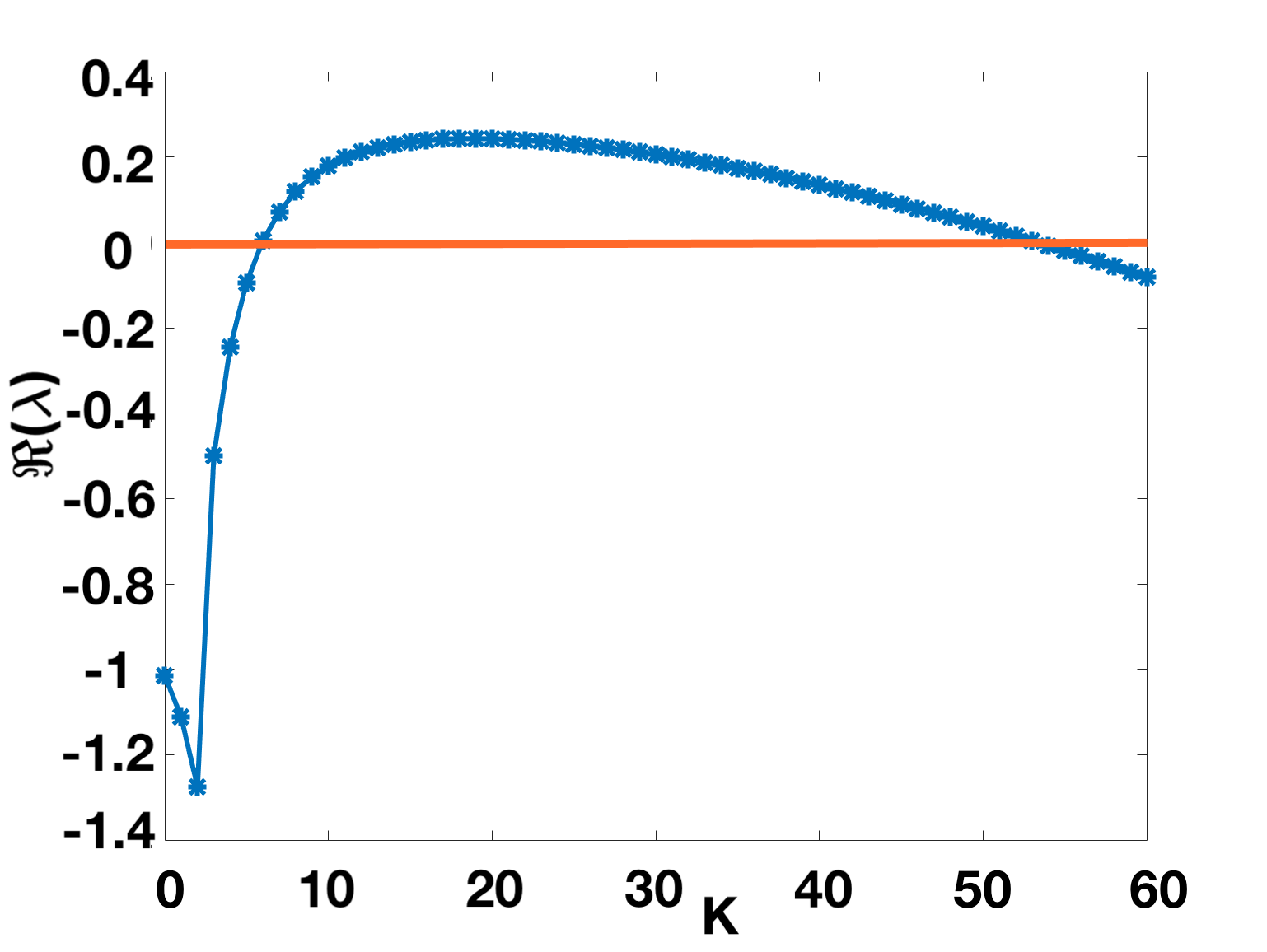}
        \caption{Dispersion curve plotted with domain size $L^2=4.5$. Curve produced by varying $k\in[0,60]$ at regular intervals of $1$.}
        \label{fig:compdisp2}
    \end{subfigure}
    \caption{Dispersion curves of the {\rd characteristic} equation given in \eqref{disp_rel} for the LI model plotted for $(a,b,\tau)=(0.4,1.8,0.2)$ and $\epsilon^2=0.001$. Discrete values of $k$ overlaid as scatter points. A larger $L^2$ results in more unstable modes $\lambda_k$ such that $\Re(\lambda_k)>0$. }
    \label{fig:compardisp}
\end{figure}

We restrict the domain size to $L^2=0.2$, as a smaller domain results in fewer unstable modes and thus less competition for the dominant mode, resulting in a better approximation of the linear theory. This finite size effect can be seen in Figure \ref{fig:compardisp}, where $\Re(\lambda_k)$ is plotted against $k$ for two different domain sizes, for a given $(a,b,\tau)$. We also use an initial small perturbation of the form of Eqn.~\eqref{firstic} for $t\in[-\tau,0]$, but vary the standard deviation, $\sigma_{IC}$, of these perturbations.

The linear theory suggests that the perturbation will be of the form $\xi(x,t)\sim A_k(t)\cos(k\pi x)$, where $k$ is the dominant mode and $A_k(t)$ denotes the corresponding Fourier coefficient at time $t$. For a given parameter set $(a,b,\epsilon^2,\tau,L^2)$, we can solve the characteristic equation \eqref{disp_rel}, and plot $\Re(\lambda_k)$ against $k$, to determine the dominant mode $k$ and the corresponding  growth rate, $\lambda_k$. When the perturbation $\xi$ has grown sufficiently, in absolute value, {\bl beyond a threshold where pattern formation is considered, we call this time $t=T^{\star}$. More specifically, the time $T^{\star}$ is the first time such that $\max_x|u(T^{\star},x)-u_\star|>u_{T^{\star}}$, where $u_{T^{\star}}$ is a given threshold. This then gives the first time such that any solution point across the whole spatial domain is large enough, in absolute difference, from the steady state. We refer to this value $T^{\star}$ as the `true' time-to-pattern. Finally, using the relation  $A_k(T^{\star})\approx A_k(0)e^{\lambda_k T^{\star}}$, we can rearrange for $T^{\star}$ and thus compute a linear {\rd theory} approximation for the `true' time-to-pattern, via
\begin{equation}\label{ttprelation}
    T^{\star}\approx \frac{1}{\lambda_k}\ln\left(\frac{A_k(T^{\star})}{A_k(0)}\right)=:T. 
\end{equation}
 The approximate time-to-pattern, $T$, can thus be determined via $\lambda_k$, from \eqref{disp_rel}, as well as $A_k(0)$ and $A_k(T^{\star})$, computed via the Fast Fourier Transform.}

\begin{figure}
    \centering
    \begin{subfigure}[t]{0.32\textwidth}
        \centering
        \includegraphics[width=5cm,height=5cm]{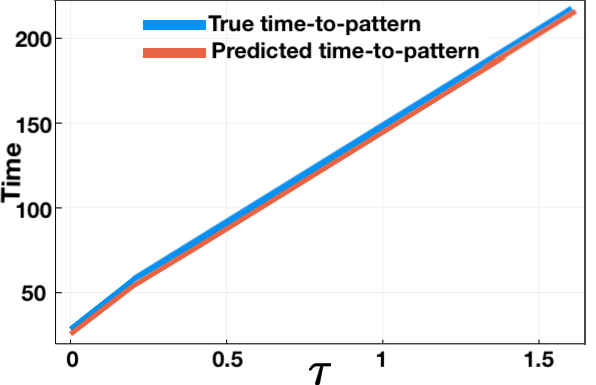}
        \caption{$(a,b)=(0.4,1.8)$.}
        \label{fig:ttp1}
    \end{subfigure}
    \hfill
    \begin{subfigure}[t]{0.32\textwidth}
        \centering
        \includegraphics[width=5cm,height=5cm]{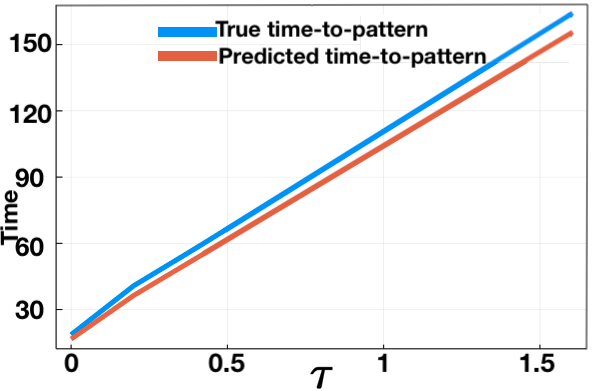}
        \caption{$(a,b)=(0.1,0.9)$.}
        \label{fig:ttp2}
    \end{subfigure}
    \hfill
    \begin{subfigure}[t]{0.32\textwidth}
        \centering
        \includegraphics[width=5cm,height=5cm]{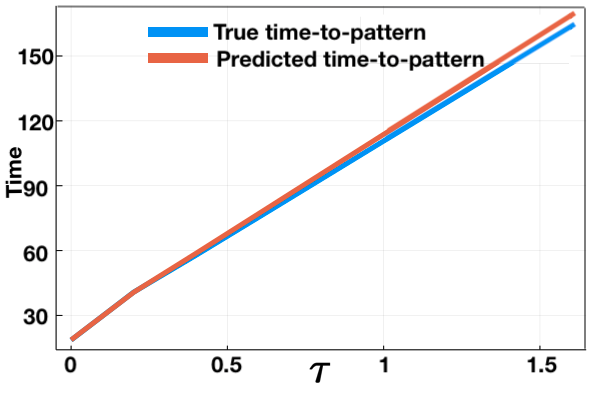}
        \caption{$(a,b)=(0.2,1.3)$.}
        \label{fig:ttp3}
    \end{subfigure}
    \caption{Predicted vs `true' time-to-pattern for numerical solutions of the LI model with fixed time delay and boundary conditions given by \eqref{neumannbc}. Initial perturbations from the steady state are taken with $\sigma_{\text{IC}}= 10^{-5}$ and a threshold of {\rd $u_{T^*}=0.1$} is used to compute the `true' time-to-pattern, $T$. The predicted time-to-pattern is computed using  \eqref{ttprelation} with $\lambda_k$ computed via \eqref{disp_rel}, for three different parameter sets, with $L^2=0.2$, $\epsilon^2=0.001$, and $\tau\in[0,1.6]$, varied at regular intervals of $0.2$.}
    \label{ttp}
\end{figure}

\begin{figure}
    \centering
    \begin{subfigure}[t]{0.45\textwidth}
        \centering
        \includegraphics[width=6cm,height=5cm]{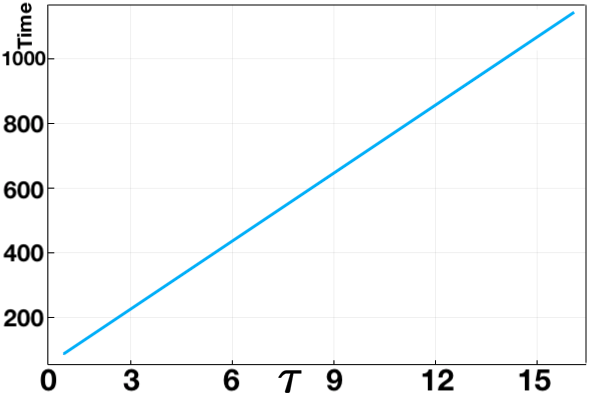}
        \caption{$(a,b)=(0.1,0.9)$}
        \label{fig:linperturb1a}
    \end{subfigure}
    \hfill
    \begin{subfigure}[t]{0.45\textwidth}
        \centering
        \includegraphics[width=6cm,height=5cm]{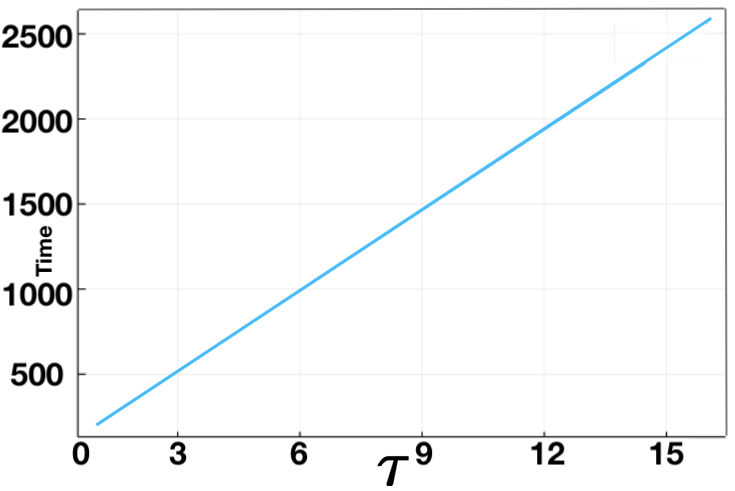}
        \caption{$(a,b)=(0.4,0.8)$}
        \label{fig:linperturb1b}
    \end{subfigure}
    \caption{`True' time-to-pattern for full numerical solutions of the fixed time delay LI model plotted against $\tau\in[1,16]$ for $\sigma_{\text{IC}}=10^{-5}$ and threshold {\rd $u_{T^*}=0.1$.} Parameters used are $\epsilon^2=0.001$ and  $L^2=4.5$.}
    \label{fig:linperturb1}
\end{figure}

We repeat  the evaluation of both the true and predicted times to pattern for varying $(a,b,\tau)$, and compare the evaluations in Figure \ref{ttp}, which shows the comparison for three different parameter sets for $(a,b)$, with the time delay varied over $\tau\in[0,1.6]$ at intervals of $0.2$. Other parameter values, and other choices of the threshold {\rd $u_{T^*}$,} gave qualitatively similar results, essentially indicating that the computed value of $\lambda_k$ for the dominant mode accurately predicts the time to pattern, which is observed to vary linearly in $\tau$  to a very good approximation\cite{sargood2022gene}. 
We remark that computing $\lambda_k$ for $\tau>1.6$ is numerically difficult; thus full numerical solutions were used to plot `true' time-to-pattern against $\tau$ on a longer timescale, for both the LI model and GM models, to verify the linear relationship between the time delay and time to pattern. In Figure \ref{fig:linperturb1} we consider 
 the time delay $\tau\in[1,16]$ at unit  intervals for the LI model, with  two different parameter sets of $(a,b)$ and plot $T$, the time taken for a perturbation to grow up to a threshold value {\rd $u_{T^*}=0.1$} from an initial perturbation of size  $\sigma_{\text{IC}}=10^{-5}$, which   once more reveals  a linearly increasing relationship between $T$ and $\tau$.

\begin{figure}
    \centering
    \begin{subfigure}[t]{0.45\textwidth}
        \centering
        \includegraphics[width=6cm,height = 5cm]{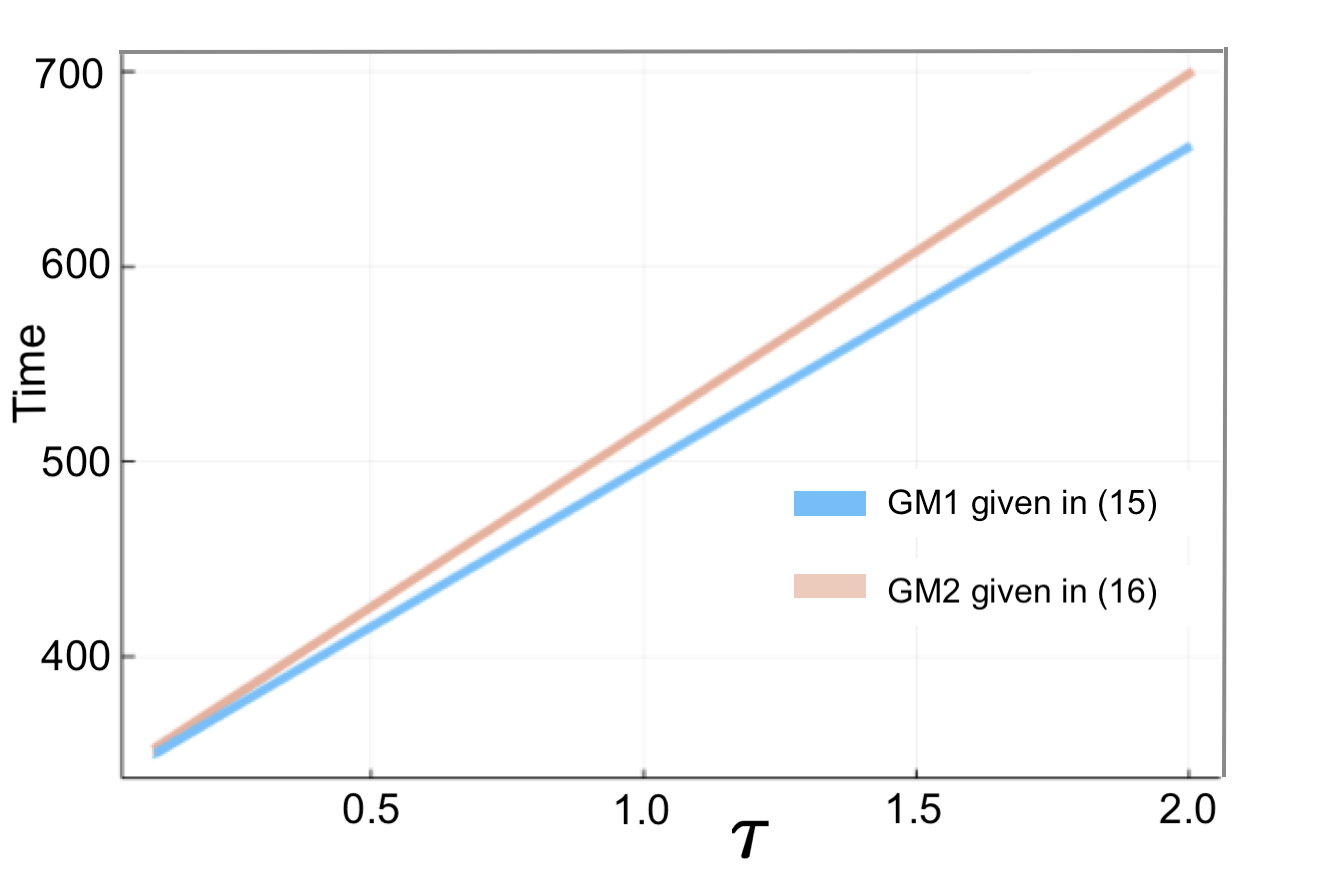}
        \caption{`True' time-to-pattern vs $\tau$ for $\tau\in[0.1,2]$ varied at intervals of $0.1$. The blue line shows results for the model in \eqref{fadai1}, and the red line for \eqref{fadai2}.}
        \label{fig:gmlin1}
    \end{subfigure}
    \hfill
    \begin{subfigure}[t]{0.45\textwidth}
        \centering
        \includegraphics[width=6cm,height = 5cm]{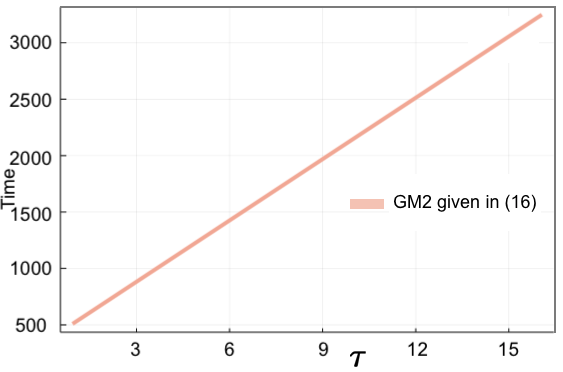}
        \caption{`True' time-to-pattern vs $\tau$ for $\tau\in[1,16]$ varied at intervals of $1$. Model given in \eqref{fadai2}.}
        \label{fig:gmlin2}
    \end{subfigure}
    \caption{`True' time-to-pattern results for the two fixed time delay GM variants given in \eqref{fadai1} and \eqref{fadai2}. Initial random perturbations given with $\sigma_{\text{IC}}=0.001$, and threshold value as {\rd $u_{T^*}=10$.} Parameters $(a,b)=(0.75,0.5)$, $\epsilon^2=0.001$ and $L^2=4.5$ are used. Boundary conditions set as in \eqref{neumannbc}.}
    \label{fig:fadlin}
\end{figure}

 For the GM$_1$ model \eqref{fadai1}, as a result of the shrinking Turing space, we only consider small $\tau\in[0.1,1]$, varied at regular intervals of $0.1$. For the GM$_2$ model \eqref{fadai2}, we consider both small $\tau\in[0.1,2]$ and larger $\tau\in[1,16]$, varied at regular intervals of $0.1$ and $1$ respectively. We set an initial perturbation from the steady state as $\sigma_{\text{IC}}=0.001$, and a threshold value of $10$ (these larger values are chosen for computational reasons, and qualitatively identical results are found for other choices).  `True' time-to-pattern results, analogous to the LI model, are presented in Figure \ref{fig:fadlin}, for both Gierer-Meinhardt models, Eqn.   \eqref{fadai1} and Eqn. \eqref{fadai2},   where a linear relationship may be observed for both models. 
Finally, we note that even 
though a linear relation is consistently observed between the time delay and time to pattern, 
there is nonetheless a difference in line slope between the different models, illustrating that kinetics can affect the sensitivity in the timing of patterning onset with the time delay.  Nonetheless, in all cases shown, the slopes are generally steep, with delays of $O(1)$ time units leading to an order-of-magnitude increase in the time-to-pattern.

\subsection{Onset of Patterning in Models with Distributed Delay}

 We have shown the linear stability analysis predicts that a continuously distributed time delay with a  symmetric or skewed Gaussian kernel has only a small (and often negligible) influence on the value of $\max_k(\Re(\lambda_k))$, regardless of the distribution properties, compared to a fixed delay model with the same mean delay. We thus look to confirm this result directly via full numerical simulations of the distributed delay cases. 

\begin{figure}
    \centering
    \begin{subfigure}[t]{0.32\textwidth}
        \centering
        \includegraphics[width=5cm,height=4.5cm]{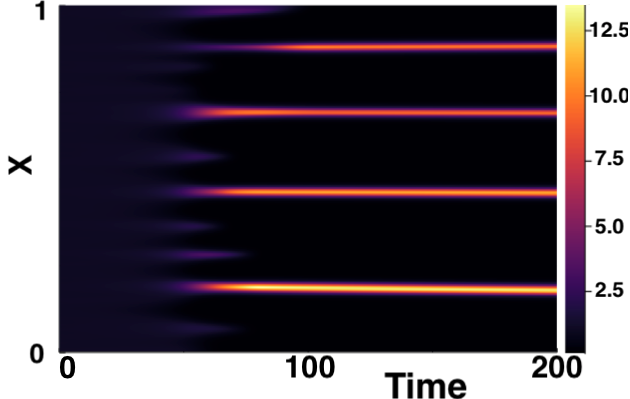}
        \caption{Fixed delay LI model.}
        
    \end{subfigure}
    \hfill
    \begin{subfigure}[t]{0.32\textwidth}
        \centering
        \includegraphics[width=5cm,height=4.5cm]{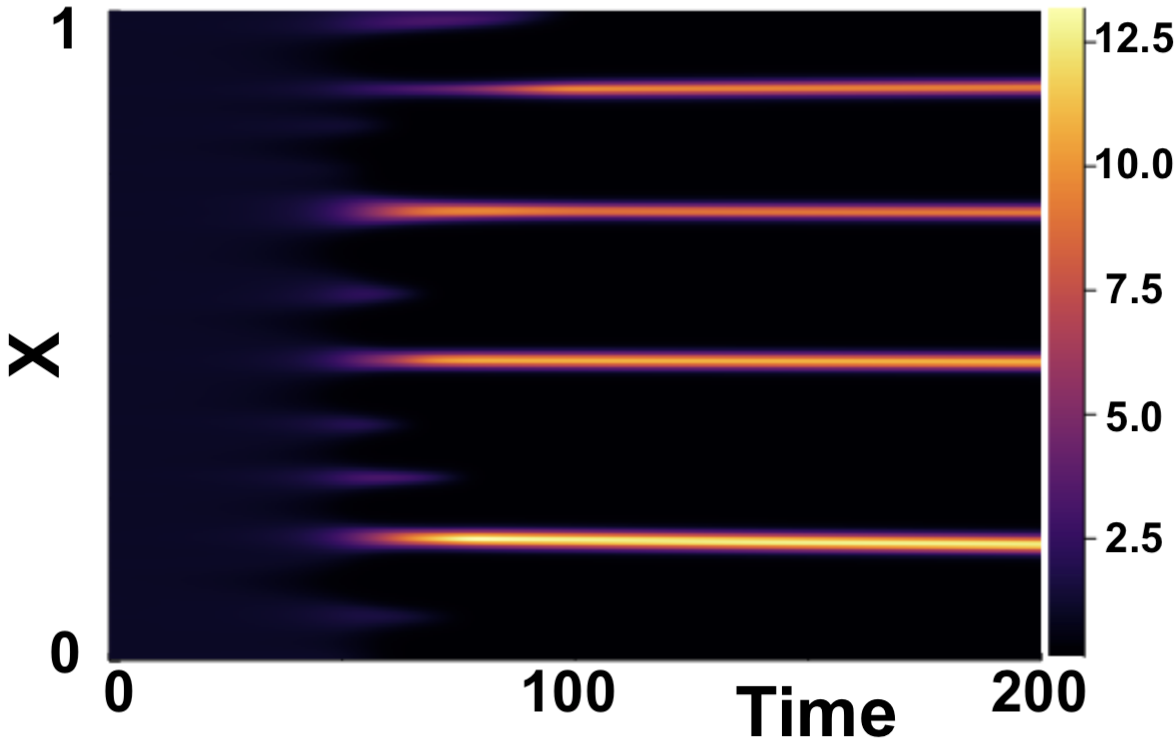}
        \caption{Distributed delay LI model with $\sigma=\sigma_{\max}\times0.99$.}
        
    \end{subfigure}
    \hfill
    \begin{subfigure}[t]{0.32\textwidth}
        \centering
        \includegraphics[width=5cm,height=4.5cm]{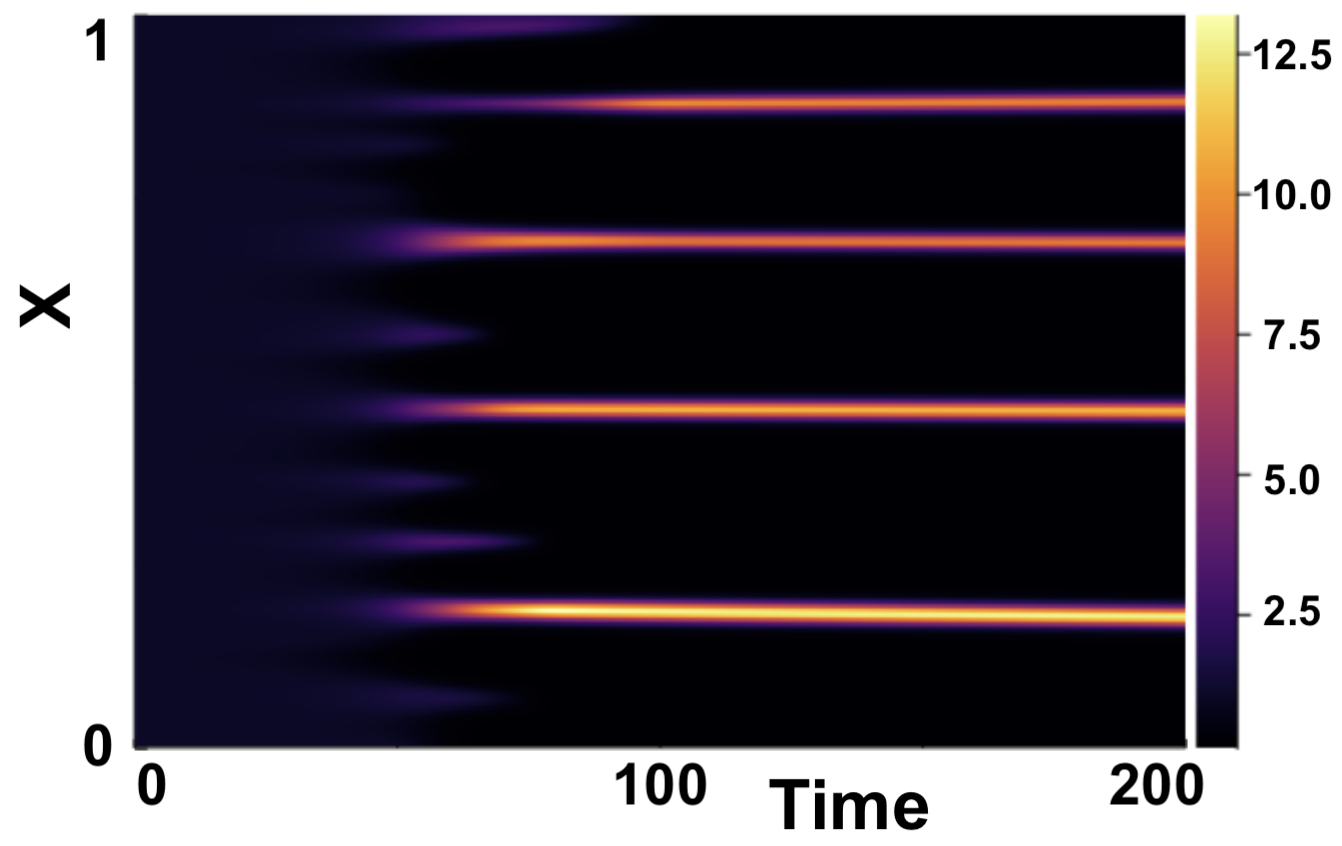}
        \caption{Distributed delay LI model with $\sigma=\sigma_{\max}\times0.1$.}
        
    \end{subfigure}
    \caption{Numerical simulations of \eqref{distmodel2} showing a comparison of the fixed vs symmetric distributed delay case for $\tau=1$. Boundary conditions are given by \eqref{neumannbc} and initial conditions by \eqref{firstic}, with parameters $(a,b)=(0.1,0.9)$, $\epsilon^2=0.001$, and $L^2=4.5$.  }
    \label{fig:distres1}
\end{figure}

\begin{figure}
    \centering
    \begin{subfigure}[t]{0.32\textwidth}
        \centering
        \includegraphics[width=5cm,height=4.5cm]{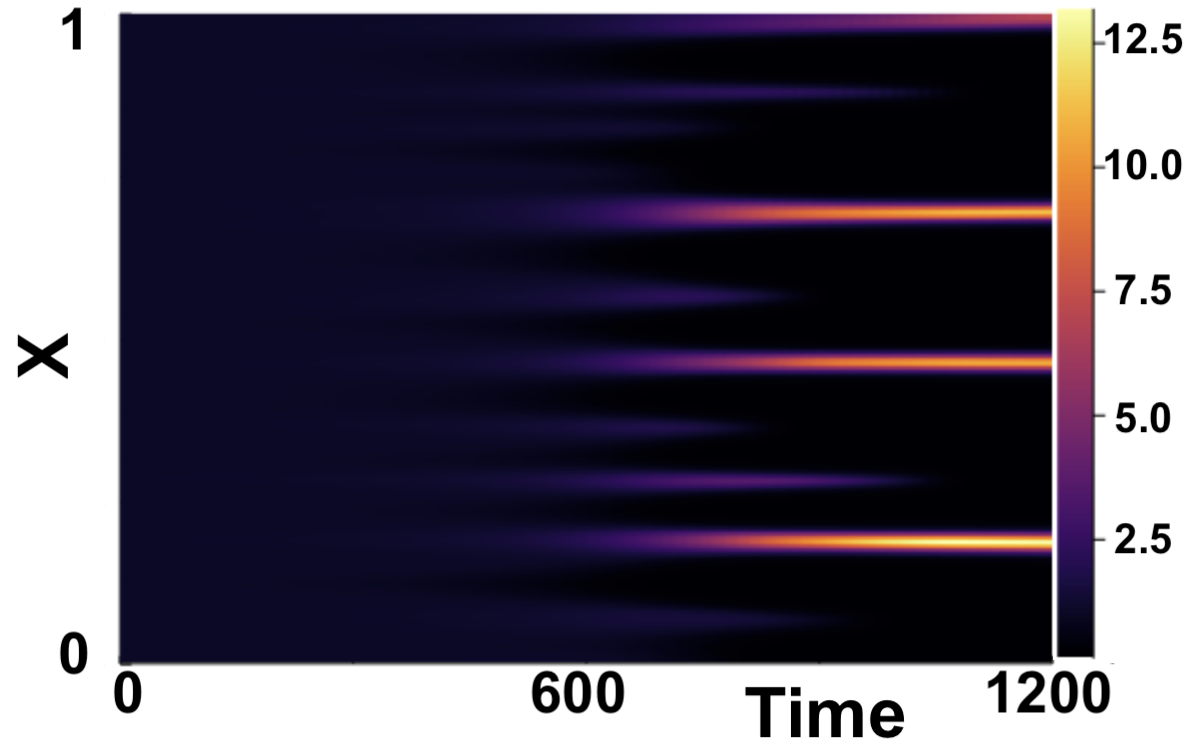}
        \caption{Fixed delay LI model.}
        
    \end{subfigure}
    \hfill
    \begin{subfigure}[t]{0.32\textwidth}
        \centering
        \includegraphics[width=5cm,height=4.5cm]{distt16sig10.png}
        \caption{Distributed delay LI model with $\sigma=\sigma_{\max}\times0.99$.}
        
    \end{subfigure}
    \hfill
    \begin{subfigure}[t]{0.32\textwidth}
        \centering
        \includegraphics[width=5cm,height=4.5cm]{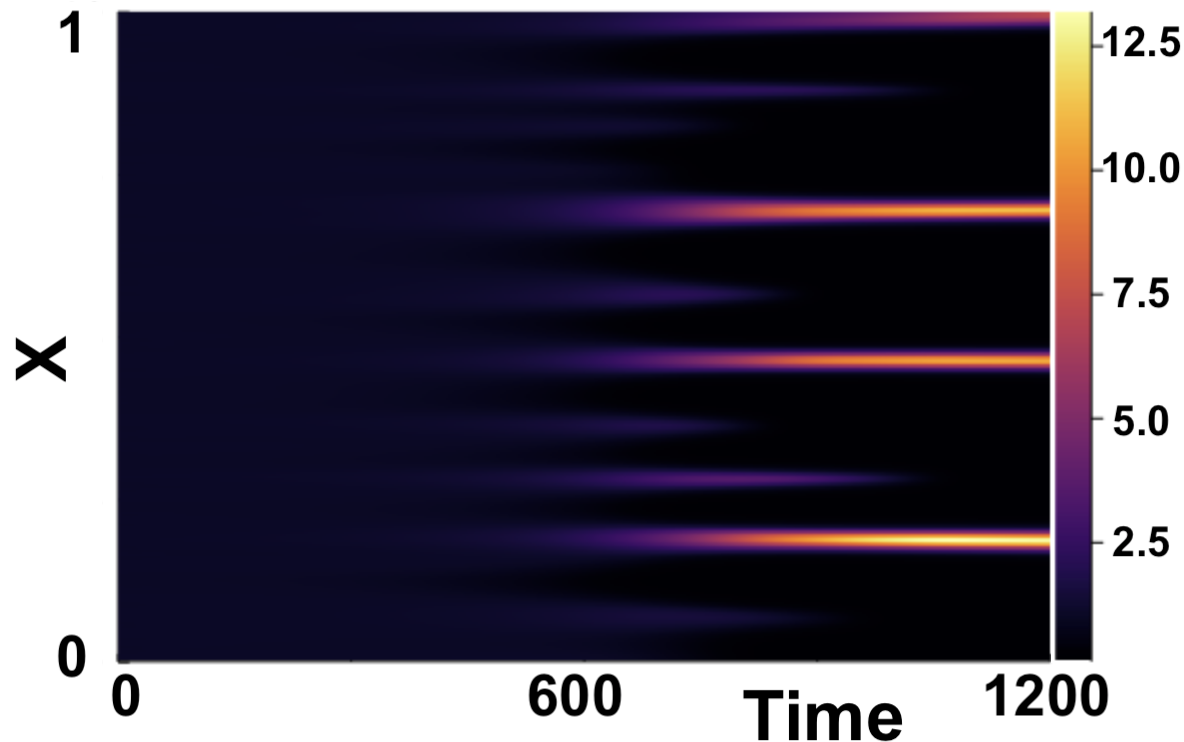}
        \caption{Distributed delay LI model with $\sigma=\sigma_{\max}\times0.1$.}
        
    \end{subfigure}
    \caption{Numerical simulations of \eqref{distmodel2} showing a comparison of the fixed vs symmetric distributed delay case for $\tau=16$. Boundary conditions are given by \eqref{neumannbc} and initial conditions by \eqref{firstic}, with parameters $(a,b)=(0.1,0.9)$, $\epsilon^2=0.001$, and $L^2=4.5$. }
    \label{fig:distres2}
\end{figure}

We first consider 
 Figures \ref{fig:distres1} and \ref{fig:distres2}, which show numerical solutions for the LI model with a symmetric Gaussian distributed delay, using $(a,b)=(0.1,0.9)$ for $\tau\in\{1,16\}$ and varying $\sigma$, and compared with the appropriate fixed delay case given an initial small perturbation of the form of Eqn.~\eqref{firstic} for $t\in[-\tau,0]$.
 The results illustrate  that the onset of patterning, and the emergent  pattern do not depend on the value of $\sigma$ used at the resolution of the plotting. 

\begin{figure}
    \centering
    \begin{subfigure}[t]{0.45\textwidth}
        \centering
        \includegraphics[width=7cm,height=4.75cm]{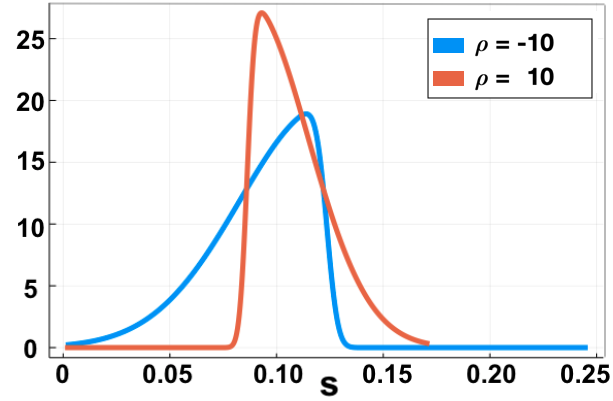}
        \caption{Probability density functions of skewed truncated Gaussian distributions, with $\rho=-10,10$ and mean $\tau=0.1$.}
        
    \end{subfigure}
    \hfill
    \begin{subfigure}[t]{0.45\textwidth}
        \centering
        \includegraphics[width=7cm,height=4.75cm]{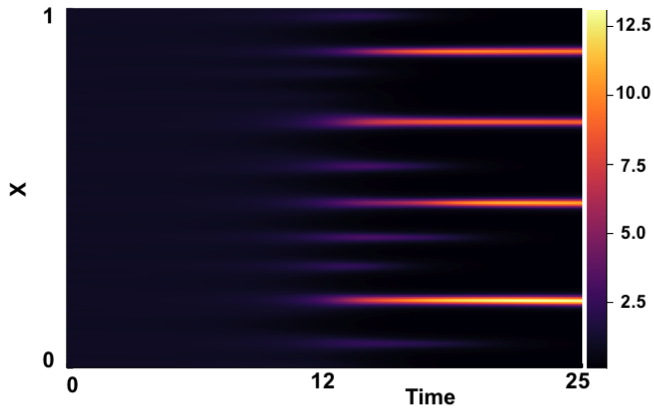}
        \caption{Fixed delay case with $\tau=0.1$.}
        
    \end{subfigure}
    \hfill
    \begin{subfigure}[t]{0.45\textwidth}
        \centering
        \includegraphics[width=7cm,height=4.75cm]{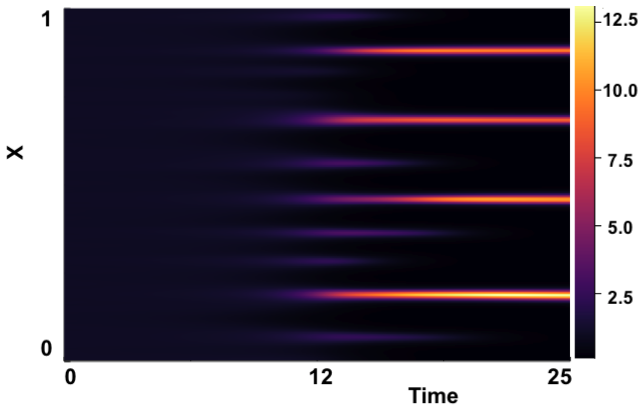}
        \caption{Numerical simulation with skewed distribution of $\rho=-10$. Distribution parameters are $\mu=0.124(3 s.f.)$ and $\omega=0.0408(3 s.f.)$.}
        \label{fig:rhom10}
    \end{subfigure}
    \hfill
    \begin{subfigure}[t]{0.45\textwidth}
        \centering
        \includegraphics[width=7cm,height=4.75cm]{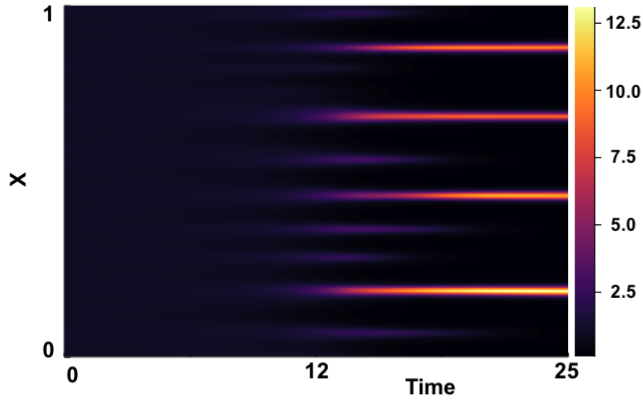}
        \caption{Numerical simulation with skewed distribution of $\rho=10$. Distribution parameters are $\mu=0.0863(3 s.f.)$ and $\omega=0.0285(3 s.f.)$.}
        \label{fig:rho10}
    \end{subfigure}
    \caption{Simulations of \eqref{distmodel2} with fixed and distributed delay with parameters $(a,b)=(0.1,0.9)$, $\epsilon^2=0.01$ and $L^2=4.5$, $\rho=-10,10$ and $\tau=0.1$. Initial conditions are given by \eqref{firstic} and boundary conditions by \eqref{neumannbc}. }
    \label{fig:linskew1}
\end{figure}

\begin{figure}
    \centering
    \begin{subfigure}[t]{0.45\textwidth}
        \centering
        \includegraphics[width=7cm,height=5cm]{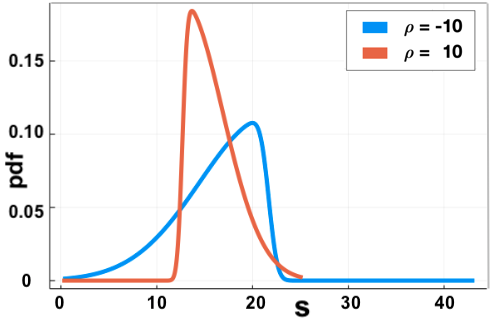}
        \caption{Probability density functions of skewed truncated Gaussian distributions, with $\rho=-10,10$, and mean $\tau=16$.}
        
    \end{subfigure}
    \hfill
    \begin{subfigure}[t]{0.45\textwidth}
        \centering
        \includegraphics[width=7cm,height=5cm]{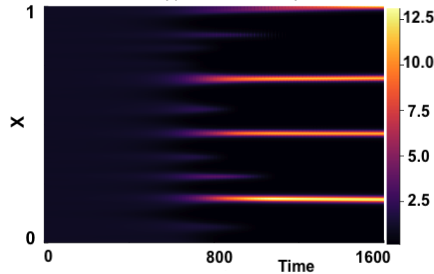}
        \caption{Fixed delay case with $\tau=16$.}
        
    \end{subfigure}
    \hfill
    \begin{subfigure}[t]{0.45\textwidth}
        \centering
        \includegraphics[width=7cm,height=5cm]{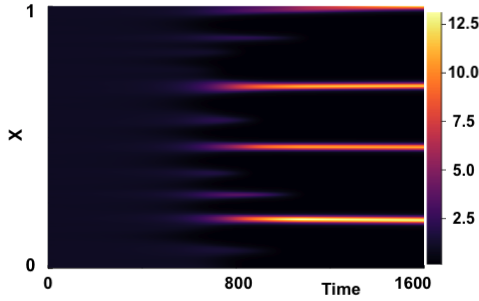}
        \caption{Numerical simulation with distribution of $\rho=-10$. Distribution parameters are $\mu=21.7(3 s.f.)$ and $\omega=7.16(3 s.f.)$.}
        
    \end{subfigure}
    \hfill
    \begin{subfigure}[t]{0.45\textwidth}
        \centering
        \includegraphics[width=7cm,height=5cm]{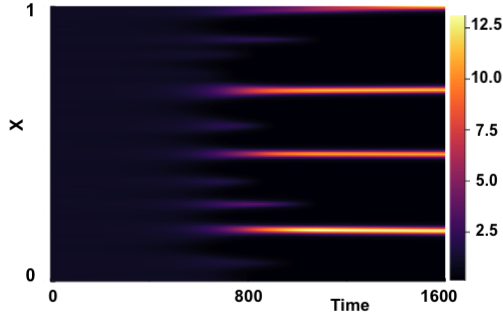}
        \caption{Numerical simulation with distribution of $\rho=10$. Distribution parameters are $\mu=12.7(3 s.f.)$ and $\omega=4.18(3 s.f.)$.}
        
    \end{subfigure}
    \caption{Simulations of \eqref{distmodel2} with fixed and distributed delay with parameters $(a,b)=(0.1,0.9)$, $\epsilon^2=0.01$ and $L^2=4.5$, $\rho=-10,10$ and $\tau=16$. Initial conditions are given by \eqref{firstic} and boundary conditions by \eqref{neumannbc}. }
    \label{fig:linskew3}
\end{figure}

We next present numerical results to show that even {\rd the skew} of an asymmetric distribution for both a small and large mean delay $\tau$, does not have a significant effect on observable  patterning results, as highlighted for the LI model by  Figure \ref{fig:linskew1} for $\tau=0.1$ and Figure \ref{fig:linskew3} for $\tau=16$. Despite a range of skews under consideration (see panel (a) in both Figures), the simulations are indistinguishable by eye to one another and to the fixed delay simulations given in panel (b) in each case. 
%Considering just the LI model, numerical simulations are shown here to  verify that modelling time delay as a Gaussian distribution (symmetric or skewed) does not typically change the results seen from that of a fixed delay model with the same mean delay, independent of the other distribution parameters used.
We finally remark that additional simulations in other parameter regimes, and for the Gierer-Meinhardt models, illustrate the same independence of the skew, with further examples can be found in \cite{sargood2022gene}. 

\subsection{Boundary and Initial Conditions} \label{BCs_ICs_Sect}

\begin{figure}
 \centering
        \includegraphics[width=16cm,height=4cm]{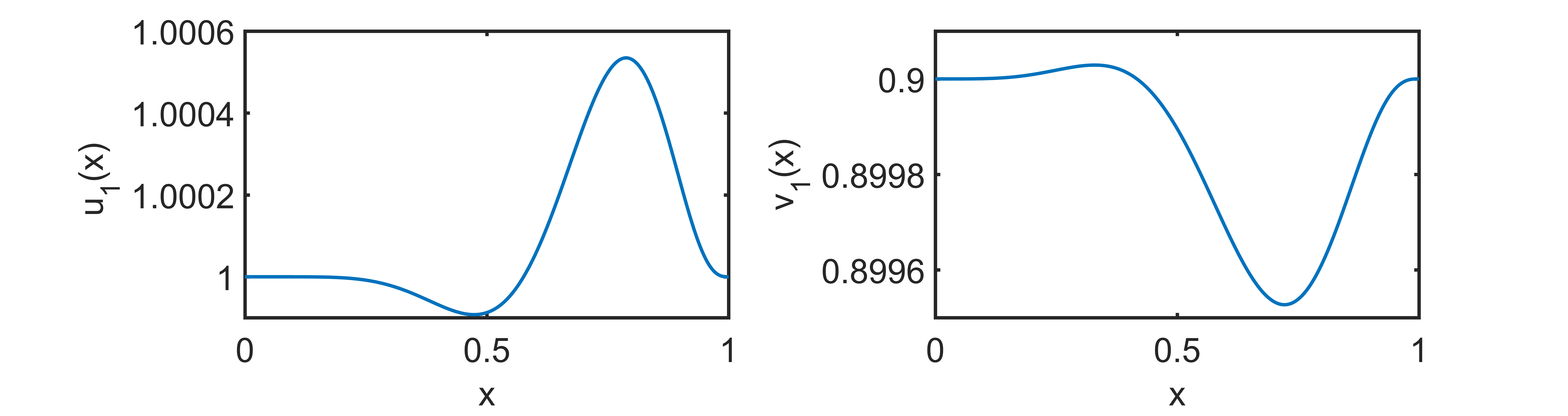}
        \caption{The functions $u_1(x)$, $v_1(x)$ used to generate the initial conditions IC$_1$; algebraic expressions for these functions are given in the Appendices of \cite{gaffmonk}.}\label{icfig}
\end{figure}
We next consider the robustness of our results under variation of the initial and boundary conditions, focusing on the LI model with a fixed delay for simplicity. We first consider the sensitivity of pattern formation in the context of a fixed time delay to varying initial conditions. 
Fixing parameters by  $$(a,b,\epsilon^2,L^2)=(0.1,0.9,0.001,4.5)$$ we proceed to consider 
three different sets of initial conditions. We denote by $\text{IC}_1$  initial conditions based on those used in \cite{gaffmonk}. With   $u_1(x)$ and $v_1(x)$ the functions depicted in Fig.~\ref{icfig}, these  conditions are given by 
\begin{equation}\label{IC1}
u(x,t) = u_1(x), ~~v(x,t) = v_1(x),
\end{equation} for $  t\in[-\tau,0]$ 
%\begin{eqnarray} \nonumber  u(x,t) = u_1(x)(1+0.0025\cos(\pi x)\cos(\pi t/(2\tau)), ~~~~v(x,t) = v_1(x)(1+0.0025\cos(\pi x)\cos(\pi t/(2\tau)), \\ \label{IC1} 
%\end{eqnarray}
%for $  t\in[-\tau,0]$ 
and represent a specific perturbation of the homogeneous steady state. We denote by $\text{IC}_2$ the initial conditions defined in Eqn.~\eqref{firstic} with $\sigma_{IC}=0.01$, and $\text{IC}_3$ denote the same initial conditions, but with $\sigma_{IC}=0.1$.   As previously,  both initial conditions were specified for $t\in[-\tau,0]$ and  a fixed random seed was set for the random variable, $r$, and hence the perturbations $r_u(x)$ and $r_v(x)$ of Eqn.~(\ref{firstic}) are the same in each simulation.

%The model parameters used match those used in \cite{gaffmonk}, with $(a,b)=(0.1,0.9)$.
 
\begin{figure}
    \centering
    \begin{subfigure}[t]{0.32\textwidth}
        \centering
        \includegraphics[width=5cm,height=4.5cm]{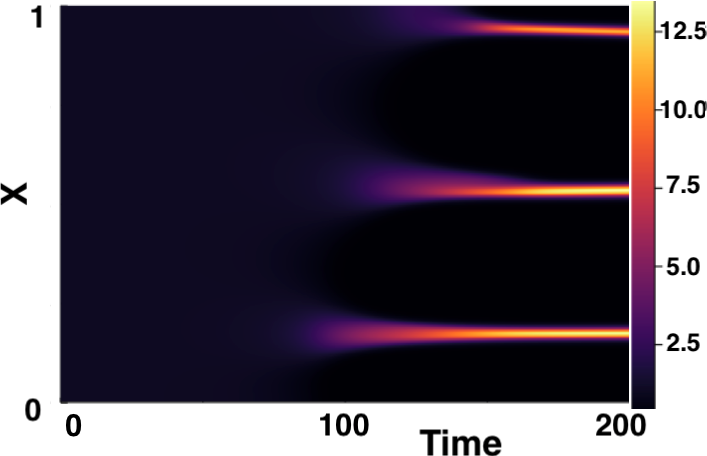}
        \caption{$\text{IC}_1$ given by Eqn.~(\ref{IC1}).}
        
    \end{subfigure}
    \hfill
    \begin{subfigure}[t]{0.32\textwidth}
        \centering
        \includegraphics[width=5cm,height=4.5cm]{ic21.png}
        \caption{$\text{IC}_2$ given by Eqn.~\eqref{firstic}, with $\sigma_{IC}=0.01$.}
        
    \end{subfigure}
    \hfill
    \begin{subfigure}[t]{0.32\textwidth}
        \centering
        \includegraphics[width=5cm,height=4.5cm]{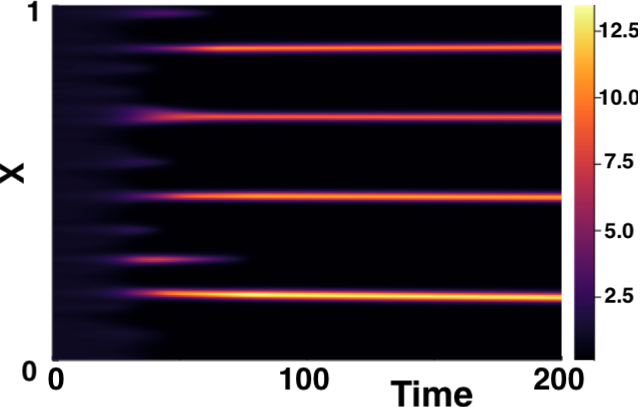}
        \caption{$\text{IC}_3$ given by Eqn.~\eqref{firstic}, with $\sigma_{IC}=0.1$.}
        
    \end{subfigure}
    \caption{Simulations of LI model \eqref{distmodel2} for varying ICs and $\tau=1$. Boundary conditions given by \eqref{neumannbc} with parameters $(a,b)=(0.1,0.9)$, $\epsilon^2=0.001$, and $L^2=4.5$.}
    \label{fig:figtau1}
\end{figure}

\begin{figure}
    \centering
    \begin{subfigure}[t]{0.32\textwidth}
        \centering
        \includegraphics[width=5cm,height=4.5cm]{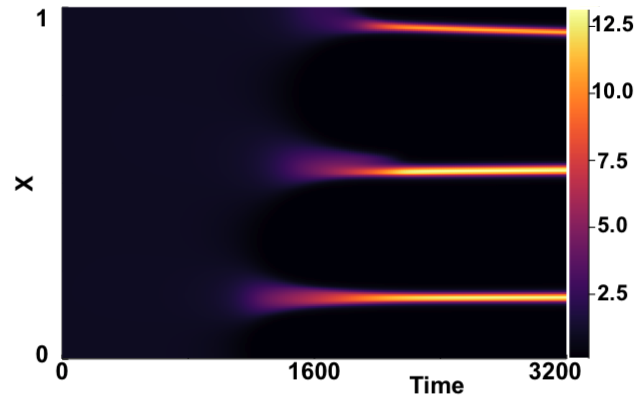}
        \caption{$\text{IC}_1$ given by Eqn.~(\ref{IC1}).}
        
    \end{subfigure}
    \hfill
    \begin{subfigure}[t]{0.32\textwidth}
        \centering
        \includegraphics[width=5cm,height=4.5cm]{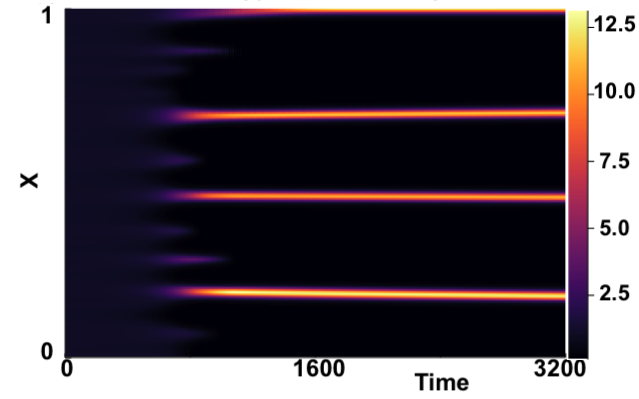}
        \caption{$\text{IC}_2$ given by equation \eqref{firstic}, with $\sigma_{IC}=0.01$.}
        
    \end{subfigure}
    \hfill
    \begin{subfigure}[t]{0.32\textwidth}
        \centering
        \includegraphics[width=5cm,height=4.5cm]{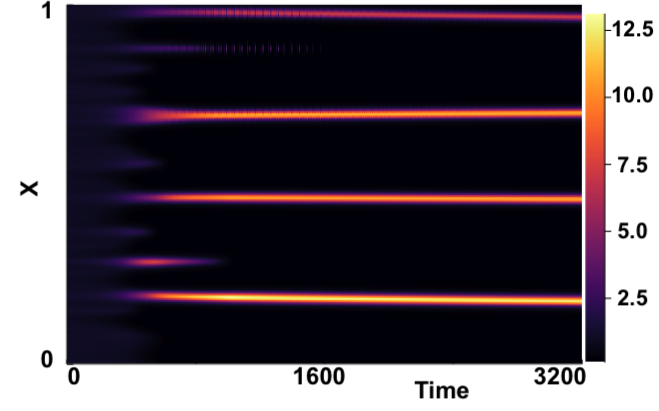}
        \caption{$\text{IC}_3$ given by Eqn.~\eqref{firstic}, with $\sigma_{IC}=0.1$.}
        
    \end{subfigure}
    \caption{Simulations of LI model \eqref{distmodel2} for varying ICs and $\tau=16$. Boundary conditions given by \eqref{neumannbc} with parameters $(a,b)=(0.1,0.9)$, $\epsilon^2=0.001$, and $L^2=4.5$.}
    \label{fig:figtau16}
\end{figure}

 Figures \ref{fig:figtau1} and \ref{fig:figtau16} show simulations for each of these initial conditions for varying fixed time delay $\tau\in\{1,16 \}$. The final pattern is somewhat sensitive to the choice of initial conditions. As one might expect, the larger $\sigma_{\text{IC}}$ used in $\text{IC}_3$, compared to that of $\text{IC}_2$, results in a faster onset of pattern formation, as does using a random perturbation rather than one along a single mode. Although the time taken until onset of patterning varies with different initial conditions, the increase in time-to-pattern with an increasing time delay is consistent independent of the initial conditions chosen. By considering the varying $x$-axis, we also note that in each case, this relationship appears to be linear, as predicted earlier. 
 
{\bl The effects of time-dependent histories were also considered. Taking initial data to be the homogeneous steady states multiplied by $1+r\sin(wt)$, with varying real $w$ and $r$ Normally distributed at each spatial point, had only transient effects with no noticeable impact on the long-time pattern structure or time to pattern results. See \cite{sargood2022gene} for examples.}

\begin{figure}
    \centering
    \begin{subfigure}[t]{0.45\textwidth}
        \centering
        \includegraphics[width=6cm,height=4.5cm]{ic21.png}
        \caption{$\tau=1$, Homogeneous Neumann boundary conditions, as  given by Eqn.~\eqref{neumannbc}.}
        
    \end{subfigure}
    \hfill
    \begin{subfigure}[t]{0.45\textwidth}
        \centering
        \includegraphics[width=6cm,height=4.5cm]{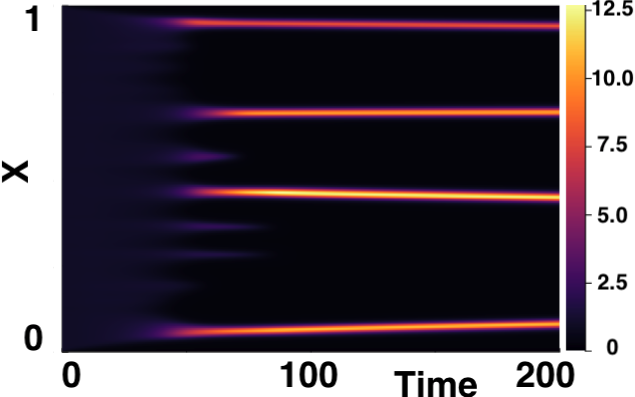}
        \caption{$\tau=1$, Mixed boundary conditions, as  given by Eqn.~\eqref{homogeneousbc}.}
        
    \end{subfigure}
     \begin{subfigure}[t]{0.45\textwidth}
        \centering
        \includegraphics[width=6cm,height=4.5cm]{ic216.png}
        \caption{$\tau=16$, Homogeneous Neumann boundary conditions, as given by Eqn.~\eqref{neumannbc}.}
        
    \end{subfigure}
    \hfill
    \begin{subfigure}[t]{0.45\textwidth}
        \centering
        \includegraphics[width=6cm,height=4.5cm]{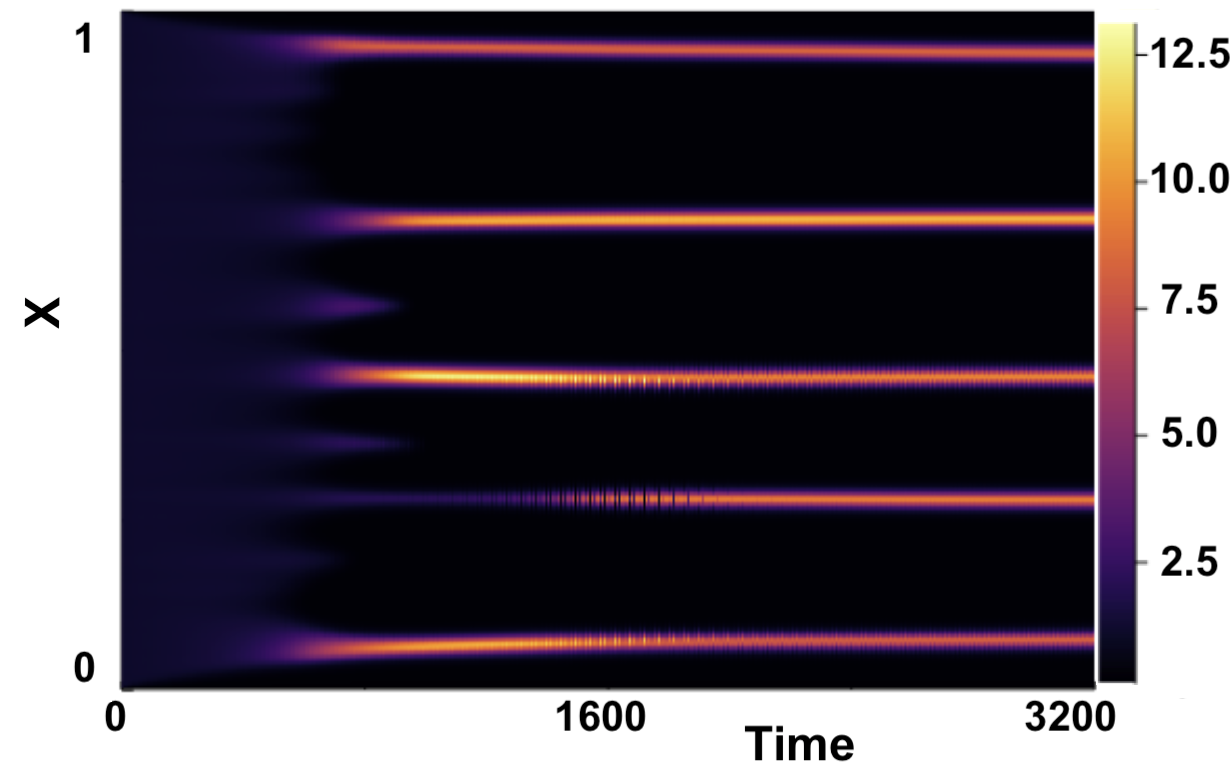}
        \caption{$\tau=16$, Mixed boundary conditions, as given by Eqn.~\eqref{homogeneousbc}.}
        
    \end{subfigure}
    \caption{Comparison of varying BCs for simulations of the LI model \eqref{distmodel2} with fixed delay and parameters  $(a,b)=(0.1,0.9)$, $\epsilon^2=0.001$,  $L^2=4.5$,  with the initial conditions specified by Eqn.~\eqref{firstic}. In the top row the time delay is fixed at $\tau=1$, and in the {\rd bottom} row $\tau=16.$ }
    \label{fig:bctau2}
\end{figure}

Finally, we consider the effect of varying boundary conditions. Motivated by the analysis in \cite{krausemixed}, homogeneous Dirichlet boundary conditions are implemented for the activator term, and homogeneous Neumann boundary conditions for the inhibitor term. Hence we set
\begin{equation}\label{homogeneousbc}
u=\frac{\partial v}{\partial x}=0 \quad x=0, 1.
\end{equation}
The results in Figures \ref{fig:bctau2} were generated using $\text{IC}_2$, with {\rd time delays of} $\tau=1$ and $\tau=16$, respectively. These show simulations generated with homogeneous Neumann conditions for both $u$ and $v$, as in the boundary conditions \eqref{neumannbc}, and those generated with the mixed conditions \eqref{homogeneousbc}. While these mixed conditions changed the kind and structure of pattern observed, as described in \cite{krausemixed} as `isolated' patterns, we do not observe a noticeable difference in the time taken for patterns to form.

\section{Discussion}\label{Discussion_Sect}

We have studied reaction-diffusion systems gene expression delays modelled as both  discrete  and  continuous distributions of delay. We used linear stability theory and systematic numerical simulation to explore the impact of varying delay distribution, as well as initial and boundary data, on the pattern formation process. Our results contribute to the clarification of  aspects of the impact of such delays from previous work, and shed light on which model features are important for consideration in biological applications.

We have provided evidence, via both linear stability and full numerical simulations, that the time until pattern onset linearly increases with increasing mean delay. This was observed across different reaction kinetics and the form of the delay, as well as initial and boundary data considered. This suggests that the impact of time delay on slowing pattern formation processes in reaction-diffusion systems is a general phenomenon which scales linearly with time delay. Although the \textit{type} of pattern seen can change with these variations (particularly for changes in boundary conditions), the increase in lag until onset of patterning as a result of time delay appears to be robust. Importantly, for the models studied here, this increase in time-to-pattern was substantial for even small and moderate delays, confirming earlier predictions in \cite{gaffmonk}. An important avenue of future work will be to understand what leads to these particular slopes, and hence refine our understanding of the plausibility of these models in the presence of gene expression time delays for pattern forming systems.

For the Schnakenberg kinetics, where fixed gene expression delays were motivated by ligand internalisation models, we have noted that increased time delays act to expand the Turing space. This effect was displayed here through the use of bifurcation diagrams, and was confirmed via numerical solutions.
Motivated by the stability analysis of spike solutions of the GM model in \cite{fadai1,fadai2}, we demonstrated the importance of the positioning of time delayed terms within a reaction-diffusion mechanism.  

We further found increasing the time delay can either  expand or contract the Turing space and this was  solely dependent on the boundary of the Turing space provided by the stability of {\rd the}  spatially homogeneous {\rd mode} in the presence of delay. 
 For the Schnakenberg ligand internalisation (LI) model this observation may be explained by first noting from Appendix A.3.1 of Gaffney $\&$ Monk \cite{gaffmonk} that a Hopf bifurcation cannot occur for a spatially inhomogeneous mode with non-zero wavenumber for this {\rd system.} Thus the growth rate $\lambda_k$ is real at a bifurcation for $k\neq 0$ and hence at the boundary of the Turing space dictated by the spatially inhomogeneous modes, we have $\lambda_k=0.$ Given the time delay {\rd only} occurs in the dispersion relation, Eqn.~\eqref{disp_rel}, via terms of the form $\exp(-\lambda_k\tau)$ these boundaries  thus must be independent of $\tau.$ Thus if the Turing space is to change for the LI model {\rd with changes in the time delay, it must be via the boundary} of the Turing space dictated by the spatially homogeneous mode, with $k=0$ and a Hopf bifurcation on the boundary, which is generally possible, as consistent with our observations. 
This further yields the question as to when the change in the Turing space with increasing time delay is solely due to the behavior of the spatially homogeneous mode, as  also observed numerically for the Gierer-Meinhardt (GM) models, in turn requiring a study of  when a Hopf bifurcation cannot occur for spatially inhomogeneous modes  given a time delay. It is an open question whether delay can induce Hopf bifurcations for inhomogeneous modes in models beyond the ones studied here. In particular, such bifurcations can occur in hyperbolic reaction-diffusion equations which can have approximately the same linearization as delayed reaction-diffusion systems for small delay \cite{ritchie_hyperbolic_2020}.

Finally, driven by the inherent stochasticity of the molecular processes underpinning gene expression \cite{raj,elowitz,mcadams,paulsson}, gene expression time delays were modelled as both a symmetric and skewed Gaussian distribution. Through linear analysis, and verified by numerical simulations, it was shown that the distribution of delay has no qualitative (and negligible quantitative) impact on solutions to these reaction-diffusion systems compared to a fixed delay model with the same mean delay. Namely, solutions of the Schnakenberg model seem to be dependent on the mean delay of the distribution used, but not on the standard deviation or skew, and thus can effectively be modelled as purely a fixed delay.

Our findings, that a distributed representation of time delay does not alleviate the increased timescales of patterning events caused with a fixed delay, reinforces earlier work indicating that such results are at odds with rapid developmental biology patterning events \cite{gaffmonk}. Of course, there are still important limitations of our simple two-species models. It is becoming increasingly clear that going beyond two-species models is crucial for representing biological reality \cite{krausenew, mainigeneral, scholes2019comprehensive}.  Hence, one potentially important avenue for further research would be to investigate the effect of time delay on Turing mechanisms encapsulating larger systems. This would aid in improving the possible applicability and similarity of Turing's models to more intricate biological dynamics.

Throughout this paper, we considered two different types of kinetics, namely those based on Schnakenberg and GM models. Our results indicate some general attributes that are common to both sets of kinetics. The first being a linearly increasing relationship between incorporated time delay and time until onset of patterning. The second being that the effect of time delay on the Turing space is only dependent on the stability of the homogeneous equilibrium in the absence of diffusion. A clear extension to these observations would be to explore these effects for different reaction-diffusion systems that can exhibit Turing patterns. Typical systems that could be examined include the Gray-Scott \cite{grayscott} or Thomas \cite{murray} models.

 The use of other forms of distribution, such as the gamma or exponential distributions could be considered, in order to verify our findings that, when onset of patterning is being considered, the only relevant modelling parameter required is the mean delay. We also only considered the problem on a one-dimensional stationary spatial domain. Previous research has been conducted on higher-dimensional spatial domains, and growing domains, with fixed delay \cite{gaffmonk,krausefixed}. Although we hypothesise that our results with a distributed delay will be consistent across variations in the spatial domain considered, there is room to explore these possibilities. Finally, we note that due to numerical limitations when using \textit{Chebfun} to find roots of the transcendental characteristic equations, the linear theory could only be tested for small-time delays. We found that, for these small-time delays, the linear theory generally provided a good approximation to the time-to-pattern, and all conclusions from the linear theory could be verified using full numerical solutions. Further work could therefore explore solving the characteristic equations derived in this paper for larger time delay values, and examine whether the linear theory still provides good approximations to the model behaviour. The development of such techniques has applicability for other non-classical dispersion relations, such as those derived in \cite{krause2020turing} which have similar numerical difficulties due to the presence of exponential functions.

\begin{acknowledgements}
In the final stages of document preparation and review, A. S. was supported by the UK Medical Research Council (G116768, SUAG/090) and The Wellcome Trust Institutional Strategic Support Fund (RG89305, RG89529).
\end{acknowledgements}

{\small 
\noindent {\bf Data Statement}
In compliance with UKRI Research Council initiatives, all code and data used to generate these Figures can be 

\vspace*{-0.8mm}\noindent  found at  
http:/\hspace*{-0.3mm}/dx.doi.org/xxx/xxx as well as at the open access repository \cite{git}.}

\printbibliography

\end{document}